\documentclass[letterpaper]{JHEP3}
\usepackage{array}

\usepackage{nicefrac}
\usepackage{amsmath}

\usepackage{amssymb,amsfonts,mathrsfs}
\usepackage{graphicx}
\usepackage{epsfig}

\usepackage{amsfonts}

\newcommand{\Tr}{\mbox{Tr}}

\def\x{\frak{x}}

\def\t{\tau}

\def\h{\eta}

\def\IC{\relax\hbox{$\inbar\kern-.3em{\rm C}$}}

\def\IC{{\bf C}}

\def\bea{\begin{eqnarray}}
\def\eea{\end{eqnarray}}
\def\be{\begin{equation}}
\def\ee{\end{equation}}
\def\ba{\begin{align}}
\def\ea{\end{align}}
\def\bse{\begin{subequations}}
\def\ese{\end{subequations}}
\def\1F1{{}_1\!F_1}
\def\2F0{{}_2\!F_0}

\def\a{\alpha}

\def\h3{$\textrm{H}_3^+$}

\def\IC{{\mathbb C}}

\def\Tr{{\rm Tr}}

\def\t{\hbox{##}}
\def\lbldef#1#2{\expandafter\gdef\csname #1\endcsname {#2}}

\def\href#1#2{#2}

\newcommand{\beq}{\begin{equation}}
\newcommand{\eeq}{\end{equation}}
\newcommand{\ber}{\begin{eqnarray}}
\newcommand{\eer}{\end{eqnarray}}

\def\be{\begin{eqnarray}}
\def\ee{\end{eqnarray}}

\def\({\left(}
\def\){\right)}
\def\[{\left[}
\def\]{\right]}
\def\<{\langle}
\def\>{\rangle}

\def\S{\mathbb S}
\def\Z{\mathbb Z}
\def\q{\frak q}
\def\z{\frak z}
\def\a{\frak a}
\def\b{\frak b}
\def\t{\frak t}

\title{\centerline{Down the rabbit hole with theories of class ${\cal S}$}}
\author{Shlomo S. Razamat\footnote{razamat@ias.edu} \; and
Brian Willett\footnote{bwillett@ias.edu}
\\
\it Institute for Advanced Study, Princeton, NJ 08540, USA\\
}

\bigskip

\abstract{
We review some of the properties of $3d$ ${\cal N}=4$ theories obtained
by dimensionally reducing theories of class ${\cal S}$. We study $3d$ partition functions, and certain limits thereof, for such theories, and the properties implied  for these by $3d$ mirror symmetry.

\

}

\begin{document}

\section{Introduction}

Whenever one computes a physical quantity and the result enjoys certain mathematical 
beauty a natural question arises whether this mathematics has a deeper physical meaning.
In recent years several beautiful physical and mathematical results have been obtained
while studying four dimensional ${\cal N}\geq 1$ supersymmetric field theories  and  three dimensional
${\cal N}\geq 2$ theories. 
In particular a great amount of information about the partition functions of such theories 
has been collected.

Let us give an example central to this note: the supersymmetric partition functions of theories
of class ${\cal S}$~\cite{Gaiotto:2009hg,Gaiotto:2009we} in $4d$. Theories of this class
are obtained by compactifying $6d$ $(2,0)$ theory on a punctured Riemann surface.
The supersymmetric partition functions on $\S^3\times \S^1$~\cite{Gadde:2011ik,Gadde:2011uv,Gaiotto:2012xa}, and
more generally on $\S^3/{\mathbb Z}_r\times \S^1$~\cite{Alday:2013rs,Razamat:2013jxa},
of class ${\cal S}$ theories of $A_{N-1}$ type 
corresponding to Riemann surface ${\cal C}_{g,s}$ of genus $g$ and having $s$ punctures  has a very robust and mathematically interesting structure.\footnote{
For a generalization of this to type D see~\cite{Lemos:2012ph,Chacaltana:2013oka}, and for discussion of E type
case see~\cite{Chacaltana:2014jba}.
} It can be written in the following form,

\be\label{maineq}
{\cal I}=\sum_{\lambda} C_\lambda^{2g-2}\;\prod_{i=1}^s \hat \psi_\lambda(a_i)\,.
\ee 
Here $a_i$ are holonomies around non trivial cycles of the geometry
 for the global symmetries of the theory associated to the punctures
of the Riemann surface.
The parameter $\lambda$ runs over the finite dimensional irreps of $A_{N-1}$ and $\hat \psi_\lambda(a)$ are orthogonal eigenfunctions of certain difference operators. 
This form of the index is not a result of a direct 
computation starting from a Lagrangian: {\it i.e.} it is not clear how to directly obtain this expression by 
localizing any path integral. In fact, the main strength of this expression is that it equally applies to theories  of class ${\cal S}$ with and without known description in terms of a Lagrangian.

One thus might wonder what is the physical problem which directly gives us~\eqref{maineq} as its answer, and what is the physical meaning of the ingredients of this equation.  Often when we have different ways to evaluate physical observables this is due to having different physical descriptions
of the same system: a duality. 
In this context, on general grounds~\cite{Gadde:2009kb}, one expects that the expression \eqref{maineq} is a result of a computation
of a correlator in two-dimensional topological quantum field theory.
However, although we can formally specify such a theory, so far it has
 not been formulated using a $2d$ Lagrangian. 

Interestingly, these question become much more straightforward when the problem is reduced down to $3d$.  The $3d$ ${\cal N}=4$ theories one obtains by dimensional reduction
 enjoy a mirror dual description in $3d$~\cite{Benini:2010uu}. 
Although we start in $4d$ with conformal theories, the $3d$ models one obtains are not conformal
and flow to an interacting fixed point in the IR. The mirror description gives an alternative
UV starting point for the flow. Moreover, although majority of the class ${\cal S}$ theories do not have
any known Lagrangian descriptions, all the mirror duals are given in terms of usual Lagrangians.
In the  $3d$ wonderland
thus many things which were either impossible or hard to imagine in $4d$ become extremely tractable.

The purpose of this note is to review some of the structure of theories of class
 ${\cal S}$: structure which becomes more transparent in $3d$.
We will take the reader on a journey starting with ${\cal N}=2$ theories in $4d$,
going down to $3d$ theories with ${\cal N}=4$ supersymmetry, and then back to $4d$ again.
On the way we will touch upon several exciting recent developments and make certain
observations. 
 By making this journey we hope to clarify 
some of the tricky points in this story and underscore some of the salient features and interconnections
between different observations.

The plan of the paper is as follows. We start in section 2 with a very brief review of theories of class ${\cal S}$ in $4d$. Then in section 3 we make the transition to $3d$. We discuss $3d$ partition functions making full use 
of ${\cal N}=4$ supersymmetry and in particular study some of the interesting limits these partition functions possess. In section 4 we discuss aspects of dimensional reduction of theories of class ${\cal S}$. Finally in section 4 we make some speculative remarks on how one might go about to understand
the $4d$ problem by knowing the $3d$ answers.

\

\section{${\cal N}=2$ class ${\cal S}$ theories in $4d$}\label{sec4d}

Before plunging into the $3d$ wonderland let us briefly review the world of $4d$ 
theories of class ${\cal S}$ and their partition functions. One constructs class
${\cal S}$ theories by taking the $6d$ $(2,0)$ theory and considering its twisted compactification
on a punctured Riemann surface ${\mathcal C}$ such that the resulting  $4d$ theory,
${\mathcal T}_{\mathcal C}$,
has ${\cal N}=2$ 
superconformal symmetry.
 At the punctures of the Riemann surface one has to specify
boundary conditions which in $4d$ translate to a choice of matter content and in particular
to a choice of the global symmetry group. The $(2,0)$ theory has an ADE classification. In 
this note we will consider theories of type $A_n$ and for simplicity we will usually take $n=1$: however, 
all of our discussion applies also for higher rank cases.

Some of the theories of class ${\cal S}$  are free, some are continuously connected 
to free theories by exactly marginal couplings, but most of them are strongly interacting SCFTs.
Performing computations for such strongly-coupled theories  a priori is an extremely hard task.
However,  certain supersymmetric $4d$ observables can be shown to be exactly equivalent
to computations in lower dimensional theories, which in general are under somewhat better
control. An example of such an observable is the $\S^4$ partition function
which is related to Liouville/Toda theory~\cite{Alday:2009aq}. 

Another set of observables which can be exactly computed are supersymmetric 
partition functions on $\S^3/{\mathbb Z}_r\times \S^1$: {\it i.e.} supersymmetric indices on
lens spaces.
It can be argued~\cite{Gaiotto:2012xa,Razamat:2013jxa,Gadde:2013dda}
 that the partition functions on $\S^3/{\mathbb Z}_r\times\S^1$
of the theory ${\cal T}_{\cal C}$ deformed by 
certain surface defects can be obtained by acting with a difference operator
on the partition function without the defect. This surface defect spans the temporal $\S^1$
and sits on one of the equators of $\S^3$ (see section 3 for more details). 
The partition functions of theories of class ${\cal S}$ then are expected
to be naturally expressible  in terms of eigenfunctions of these operators, $\hat \psi_\lambda(z)$, as
in equation~\eqref{maineq}.  One way to argue for this is by studying analytical properties
of the partition functions. The arguments of~\cite{Gaiotto:2012xa}  lead to the conclusion
that residues at certain poles of the partition function of a theory 
obtained from ${\mathcal T}_{\mathcal C}$ by coupling to it a bi-fundamental hypermultiplet
give the partition function in presence of a surface defect. Computing such a residue is 
equivalent to an action of a difference operator. 

Mathematically these statements translate for $\hat\psi_\lambda(z)$ to the following,
\be
&& {\frak S}_{z^*}(z) \cdot \hat \psi_\lambda(z) = {\cal E}_{z^*}^\lambda \,
\hat \psi_\lambda(z)\,.
\ee
The difference operators are labeled by a pole $z^*$  in flavor fugacity which also labels
a certain choice of the surface defect.  The difference operators ${\frak S}_{z^*}(z)$ for $\S^3\times \S^1$ are given by polynomials of  (properly conjugated) Ruijsenaars-Scnheider (RS) Hamiltonians~\cite{Gaiotto:2012xa}.\footnote{Integrable models of RS type are ubiquitous when
studying
${\cal N}=2$ supersymmetric field theories, see {\it e.g.}~\cite{Gorsky:1993pe,Gorsky:1993dq,Gorsky:1994dj,Nekrasov:2009rc}.
} For $\S^3/{\mathbb Z_r}\times \S^1$
these are certain matrix valued generalizations of the latter~\cite{Razamat:2013jxa}.
Moreover, since one obtains these difference operators by studying residues of the partition functions
the equality~\eqref{maineq} implies that
\be\label{residuesQ}
&&Res_{z\to z^*}\;\hat \psi_\lambda(z) =
{\cal E}_{z^*}^\lambda \, .
\ee
 Note that here we have to be very careful with all the normalizations for the equations
to be consistent.

\

Let us give two examples of difference operators introducing surface defects which will be relevant for this paper. First, the basic operator 
introducing a surface defect in $\S^3\times \S^1$ 
computation (for $A_1$ theories) spanning one of the equators of $\S^3$ and the $\S^1$ is given
 by~\cite{Gaiotto:2012xa}
\be\label{S3Op1}
{\frak S}_{z^*=t^{\frac12}q^{\frac12}}\,\cdot\, {\cal I}(b) \sim
\frac{\theta(\frac{t}q b^{-2};p)}{\theta(b^2;p)}\;{\cal I}(q^{\frac12}b)+
\frac{\theta(\frac{t}q b^{2};p)}{\theta(b^{-2};p)}\;{\cal I}(q^{-\frac12}\,b)\,.
\ee The meaning of the parameters appearing in this formula will be explained in the next section.
Here we used the theta-function $$\theta(z;q)=\prod_{\ell=0}^\infty(1-z\,q^\ell)(1-z^{-1}q^{\ell+1})\,.$$ The operator introducing a surface defect on the other equator of $\S^3$ is obtained by exchanging 
$p$ and $q$ in the above formula.  The joint eigenfunctions of these operators are closely related
to elliptic generalizations of Macdonald polynomials.

The second example is an operator introducing a pair of surface defects, one on each equator of $\S^3$,
of the lens space partition function. This operator is given by~\cite{Razamat:2013jxa}
\be\label{LensOp1}
&&{\frak S}^{(r)}_{z^*=t^{\frac12}(pq)^{\frac{1}2}}\,\cdot\, {\cal I}(b,m) \sim
\frac{1}{\theta(q^{2m}b^{-2};q^r)\theta(p^{2m}b^2;p^r)}\times\\
&&\quad\left(\theta(q^{2m}\frac{t}{pq}b^{-2};q^r)\theta(p^{2m}\frac{pq}{t}b^2;p^r)
{\cal I}((pq)^{\frac12}b,m)+\right.\nonumber\\
&&\qquad+\theta(q^{2m}\frac{pq}{t}b^{-2};q^r)\theta(p^{2m}\frac{t}{pq}b^2;p^r){\cal I}((pq)^{-\frac12}b,m)+
\nonumber\\
&&\qquad\quad
+\left(\frac{pq}{t}\right)^{\frac{2+4m-r}{r}}\theta(q^{2m}\frac{pq}tb^{-2};q^r)\theta(p^{2m}\frac{pq}{t}b^2;p^r){\cal I}((p/q)^{-\frac12}b,m+1)+\nonumber\\
&&\qquad\qquad\left.+\left(\frac{pq}{t}\right)^{\frac{2-4m+r}{r}}\theta(q^{2m}\frac{t}{pq}b^{-2};q^r)\theta(p^{2m}\frac t{pq}b^2;p^r) {\cal I}((p/q)^{\frac12}b,m-1)\right)\,.\nonumber
\ee For $r=1$ this operator is proportional to
 ${\frak S}_{z^*=t^{\frac12}q^{\frac12}}{\frak S}_{z^*=t^{\frac12}p^{\frac12}}$. However, for $r>1$
\eqref{LensOp1} is the basic operator surviving the $\mathbb Z_r$ projection. It is hard 
to find the explicit spectrum of eigenfunctions of these operators. However, in what follows we will
encounter the $3d$ versions of~\eqref{S3Op1} and of \eqref{LensOp1} and will discuss a very explicit and physical set  of their eigenfunctions. 

\

\

\section{Brief review of ${\cal N}=4$ $3d$ generalities}

Let us first recall some of the basic properties of the
three dimensional gauge theories with $\mathcal{N}=4$ supersymmetry.     The fields and UV actions of these theories can be obtained by dimensionally reducing those of $\mathcal{N}=1$ $6d$ gauge theories or $\mathcal{N}=2$ $4d$ gauge theories.  Recall the $6d$ theory has an $SU(2)$ R-symmetry - upon dimensional reduction, one obtains an additional $SU(2)$ factor in the R-symmetry group from rotations in the compactified dimensions.  The full R-symmetry group in $3d$ is thus $SU(2)_H \times SU(2)_C$.  The supercharges are Majorana spinors in Minkowski signature, and come in a representation $(2,2)$ of the R-symmetry.

The fields are organized into vector multiplets and hypermultiplets, along with their twisted counterparts.  Let us write the field content of these multiplets in a notation where the R-symmetry transformation properties are explicit by introducing indices $A,B,...=1,2$ for $SU(2)_C$ and $M,N,...$ for $SU(2)_H$.  For the vector multiplet, the dynamical fields are in the adjoint representation of the gauge group, and can be written as:\footnote{Here the fields are real in the sense that they satisfies reality conditions ${\Lambda^{A,M}}^\dagger = \epsilon^{AB} \epsilon^{MN} \Lambda_{B,N}$ and ${\Phi^{[AB]}}^\dagger = \epsilon^{AC} \epsilon^{BD} \Phi_{[BD]}$.}

\be \mbox{gauge field:   }A_\mu, \;\;\; \mbox{real gaugino:   }\Lambda_{A,M}, \;\;\;\mbox{real scalars:   } \Phi_{[AB]}\,.\ee
Here the bracket denotes symmetrization, so that $\Phi_{[AB]}$ is an $SU(2)_C$ triplet of scalars in the vector multiplet.  The transformations for the vector multiplet can be closed off-shell if we introduce real auxiliary scalars $D_{[MN]}$, transforming in a triplet of $SU(2)_H$.    A supersymmetric Yang-Mills action takes the form

\be
&& S_g =
\int d^3 x \frac{1}{g^2} \mbox{Tr} \left( F_{\mu \nu} F^{\mu \nu} +\right.\\
&&\qquad\left.+ D_\mu \Phi_{[AB]} D^\mu \Phi^{[AB]} + i \Lambda_{A,M} 
D \!\!\!\!/\; \Lambda^{A,M} + \epsilon^{MN} \Lambda_{A,M} \Lambda_{B,N} \Phi^{[AB]}+ D_{[MN]} D^{[MN]}  \right)\,.\nonumber
\ee
Note that in $3d$ one can also include a Chern-Simons kinetic term for the gauge field.  This is, however, incompatible with the ${\mathcal{N}=4}$ supersymmetry transformations preserving the action above.
There are also special examples with enhanced supersymmetry, such as the ABJM theory, where one considers Chern-Simons actions with no Yang-Mills term.  We will not consider such theories in this paper.

The supersymmetry transformations of the hypermultiplet cannot be closed off-shell for the full ${\mathcal N}=4$ superalgebra.  For now we will be content with working with the on-shell fields:

\be \mbox{complex scalar:  }Q_{M} \;\;\; \mbox{complex fermion:   }\Psi_{A} \,.\ee
These can be taken in a representation $R$ of the gauge group and coupled to a vector multiplet, with the following action:

\be 
&& \qquad\qquad S_m = \int d^3 x ( D_\mu {Q^M}^\dagger D^\mu Q_M + i {\Psi^A}^\dagger D \!\!\!\!/\; \Psi_A + \\
&& + {Q^M}^\dagger \Phi^{[AB]} \Phi_{[AB]}Q_M  + {\Psi^A}^\dagger \Phi_{[AB]} \Psi^{B} + {\Psi^A}^\dagger \Lambda_{A,M} Q^M +{Q^M}^{\dagger} \Lambda_{A,M}{\Psi^A}  + {Q^M}^\dagger D_{[MN]} Q^N )\,.\nonumber
\ee

In addition to the field content above, there are twisted vector and hyper multiplets \cite{Brooks:1994nn,Kapustin:1999ha}, which are as above, except with their $SU(2)_C$ and $SU(2)_H$ transformation properties exchanged.  Then we can couple twisted hypermultiplets to twisted vector multiplets, although not to ordinary vector multiplets.  We can couple a twisted vector multiplet to an ordinary vector multiplet, provided one of them is abelian, by a BF term (assuming the twisted vector multiplet is abelian, and writing its fields with a prime):

\be S_{BF} = \int d^3 x \mbox{Tr}(A' \wedge F + \Phi_{[AB]} {D'}^{[AB]} + D_{[MN]} {\Phi'}^{[MN]}+ \Lambda_{A,M} {\Lambda'}^{M,A} )\,,
\ee
where $F$ is the field strength for the ordinary vector multiplet.

The moduli spaces of these theories can roughly be split into a Higgs branch, where the hypermultiplet scalars get VEVs, and a Coulomb branch, where the scalars in the vector multiplet get VEVs.  These are both hyper-Kahler manifolds, and the former does not receive any quantum corrections, so can be computed exactly by studying the $D$-term equations in the UV theory.  From the transformation properties of the corresponding scalars, we can see the Higgs branch is acted on by the $SU(2)_H$ symmetry and the Coulomb branch by the $SU(2)_C$, which explains their names.  There may also be mixed branches where both kinds of scalars get VEVs.

Finally, we can also add mass and Fayet-Iliopolos (FI) terms to the action.  The mass (FI) parameters live in background vector (twisted vector) multiplets, and are associated to global symmetry groups.   First consider mass terms.  These are associated to a global flavor symmetry group $G_H$, which acts on the Higgs branch scalars of the theory \cite{Gaiotto:2013bwa}.  The mass term can be obtained by giving an expectation value to the scalar $\Phi_{[AB]}$ in a background vector multiplet coupled to this symmetry.  Thus they come in an $SU(2)_C$ triplet, $M_{[AB]}$, and enter the action as terms:

\be S_{mass} = \int d^3 x( {Q^M}^\dagger M^{[AB]} M_{[AB]}Q_M  + {\Psi^A}^\dagger M_{[AB]} \Psi^{B} )\,.\ee
In $\mathcal{N}=2$ notation, the $M^{[AB]}$ decompose as a real mass and a complex (superpotential) mass.

The FI term, on the other hand, can be thought of as living in a background twisted vector multiplet, and is associated to a $U(1)$ factor of the gauge group.  It couples via a BF term:

\be S_{FI} = \int d^3 x \mbox{Tr}( D_{[MN]} {\zeta}^{[MN]} ) \,.\ee
These are also associated to a global symmetry group, which we call $G_C$, whose maximal torus is the set of $U(1)_J$ topological symmetries, with current $J_i=\star Tr F_i$, which appear for each $U(1)$ factor in the gauge group.  This symmetry is sometimes enhanced in the IR to a larger, nonabelian symmetry group.  Then the FI terms arise by coupling the twisted vector multiplet to this symmetry group and turning on a VEV for the scalar, as for the mass terms above.

Three dimensional mirror symmetry \cite{Intriligator:1996ex,deBoer:1996mp} is a class of dualities between three dimensional $\mathcal{N}=4$ theories, which is characterized by the fact that the two R-symmetry factors, $SU(2)_H$ and $SU(2)_C$, are exchanged.  We will, however, use a notation where the same R-symmetry group acts on both theories, so that, if ordinary vector and hypermultiplets appear on one side of the duality, then twisted vector and hypermultiplets appear on the other.  As a consequence, the Higgs branch of one theory maps to the Coulomb branch of the other, and mass and FI terms are exchanged.  The simplest example is the duality between SQED with a single charge $1$ hypermultiplet on one side and a free twisted hypermultiplet on the other.  We will consider several examples of mirror symmetries, and see explicitly in index computations how the two $SU(2)$ R-symmetry factors are exchanged.

\subsection{${\cal N}=4$ $3d$ partition functions}\label{defsec}

We would like to study $3d$ partition functions of ${\cal N}=4$ theories.  In particular we are interested in properties which are evident when exploring the extended supersymmetry of these theories. 

We will be mainly interested in $3d$ $\mathcal{N}=4$ theories obtained by dimensional reduction from $4d$ $\mathcal{N}=2$ theories (see, eg \cite{Seiberg:1996nz}).  We will thus define $3d$ ${\cal N}=4$ supersymmetric partition functions by dimensionally reducing the $4d$ ${\cal N}=2$ partition functions.\footnote{These partition functions can, of course, also be defined intrinsically in three dimensions.  When we consider partition functions of $\mathcal{N}=4$ $3d$ theories which do not reduce from four dimensional partition function, for example, those whose $4d$ lift would be non-conformal, we should use the intrinsic $3d$ definitions. (see, eg, \cite{Kim:2009wb,Imamura:2011su,Kapustin:2009kz,Jafferis:2010un})}  In this note we will be in particular interested in partition functions on $\S^3$ and $\S^2\times \S^1$, both of which can be understood 
from reducing partition functions on $\S^3/{{\mathbb Z}_r}\times \S^1$~\cite{Benini:2011nc}. Let us start then by defining the latter $4d$ partition function,
\be
{\cal I}=\Tr_{\S^3/\Z_r}(-1)^F\, p^{j_2+j_1-\hat r}\,q^{j_2-j_1-\hat r} \, t^{\hat r+R}\,
e^{-\beta\,(E-2j_2-2R+\hat r)}\,.
\ee Here $j_2$ and $j_1$ are the Cartans of $SU(2)_{j_1}\times SU(2)_{j_2}$ isometry 
of $\S^3$; the charges $\hat r$ and $R$ are the Cartans of the R-symmetry $SU(2)_R\times U(1)_{\hat r}$;
and $E$ is the energy in the radial quantization. 
The space $\S^3/\Z_r$ is defined as follows. We parameterize the $\S^3$
as 
\be
(z_1,z_2),\qquad |z_1|^2+|z_2|^2=1\,.
\ee So the two equators are $z_1=0$ and $z_2=0$. The Hopf fibration is given by the map 
of $\S^3$ to $\S^2$,
\be
(z_1,z_2)\qquad\to\qquad (2z_1z_2,|z_1|^2-|z_2|^2)\,.
\ee The Hopf fiber is parameterized by phase $\lambda$
\be
(\lambda z_1,\lambda^{-1}z_2),\qquad |\lambda|=1\,.
\ee Then $\S^3/\Z_r$ is defined by the following identifications
\be
(z_1,\,z_2)\sim(\exp\left(\frac{2\pi i }{r}\right)\, z_1,\exp\left(-\frac{2\pi i }{r}\right)\,z_2)\,.
\ee

The supersymmetric configurations of a $U(1)$ gauge field are labeled by a holonomy $z$ around the $\S^1$ cycle and $e^{2 \pi i m/r}$ around the (non-contractible) image of the Hopf fiber.  A hypermultiplet in this background has partition function:
\be
\label{lensdef}
 \mathcal{I}^{(4d)}_H(z,m;p,q,t) =
 ((pq)^\frac{1}{2} t)^{m(r-m)/r} \Gamma_e( t q^{r \mp m} z^{\pm 1};q^r,pq) \Gamma_e(t p^{\pm m} z^{\pm 1};p^r, pq) \,,
\ee where
\be
\Gamma_e(z;p,q)=\prod_{i,j=0}^\infty \frac{1-p^{i+1}q^{j+1}z^{-1}}{1-p^i q^j z}\,,
\ee
is the elliptic gamma function.  It will be convenient to redefine the fugacity $z$ as $ z \rightarrow z (p/q)^{-m/2} $, so that the contribution of the hyper becomes:\footnote{This redefinition amounts to measuring the momentum ($j_{1,2}$) of the fields with the gauge-covariant derivatives, $D_{1,2}$, rather than the non-covariant $\partial_{1,2}$.}
\be
\label{lensredef}
&& \mathcal{I}_H(z,m;p,q,t) \to \\
&&\qquad\qquad((pq)^\frac{1}{2} t)^{m(r-m)/r} \Gamma_e( t (p q)^{(r \mp m)/2} (p/q)^{r/2} z^{\pm 1};q^r,pq) \Gamma_e(t (p q)^{\pm m/2} z^{\pm 1};p^r, pq)\,. \nonumber 
\ee

Before moving on to the examples in $3d$, let us make a general comment about partition functions of $\mathcal{N}=4$ theories.  When we perform localization of these theories, we must choose a priveleged $\mathcal{N}=2$ subalgebra and corresponding $U(1)$ R-charge.  For example, we will pick an $\mathcal{N}=2$ R-charge, which we take as a $J_3^H + J_3^C$, where $J_3^i$ denotes the Cartan generators of the two $SU(2)$ $\mathcal{N}=4$ R-symmetry factors.  The other combination, $J_3^H - J_3^C$, which we will call $J_t$, appears as a flavor symmetry from the $\mathcal{N}=2$ point of view.  Note that, under mirror symmetry, where the two $SU(2)$ factors are exchanged, $J_t$ will be exchanged with $-J_t$, and we will indeed observe this explicitly in examples below.  From the point of view of this subalgebra, the component of $M_{[AB]}$ which is fixed by the Cartan of $SU(2)_C$ looks like a real mass parameter from the $\mathcal{N}=2$ point of view, while the others look like superpotential masses, and break this choice of $U(1)$ R-symmetry.  We will consider turning on only the former, real mass parameters.  Similarly, we will only turn on a single component of the FI term $\zeta_{[MN]}$, the $\mathcal{N}=2$ FI term.  Once again, such parameters will generically be exchanged by mirror symmetry.  We may also turn on a real mass for the $J_t$ symmetry defined above, although this will break us down to $\mathcal{N}=2$ supersymmetry.  On the curved manifolds which we will place these theories on, these statements about real masses map to analogous statements about which background BPS vector multiplet configurations we can turn on, whose VEVs will give the parameters on which the partition functions depend.

\

\subsection*{The $\S^2\times \S^1$ partition function}

One obtains the  $3d$ index by  sending 
the parameter $r$ to infinity. The charge $j_1$ counts the momentum on the Hopf fiber which shrinks to zero size in this limit. We set 
\be
\label{indred}
p\to {\frak q}^{1/2}y,\qquad\qquad
q\to {\frak q}^{1/2}y^{-1},
\qquad\qquad t\to {\frak t}\,{\frak q}^{1/2}\,,
\ee 
and make the following map between the $4d$ and $3d$ charges,
\be
R=R_H,\qquad\qquad \hat r=-R_C\,.
\ee Where $SO(4)\sim SU(2)_{H}\times SU(2)_{C}$  is the ${\cal N}=4$ R-symmetry in $3d$.
The $3d$ index thus obtained takes the form

\be\label{index3d}
{\cal I}=\Tr_{\S^2}(-1)^F\, \q^{j_2+\frac12(R_H+R_C)}\, \t^{R_H-R_C}\,e^{-2\beta(\widetilde E-R_H-R_C-j_2)}\,.
\ee Note that with the redefinition which gives (\ref{lensredef}), the fugacity $y=\sqrt{p/q}$ decouples in the $r \rightarrow \infty$ limit from the index of the hypermultiplet, and one can check that  it also does so for the indices of the vector fields: thus the $3d$ expressions do not depend on $y$.
 This is the ${\cal N}=4$ index we will compute, with possible further refinement with
fugacities and background magnetic fluxes for flavor symmetries. The $3d$ conformal dimension
$\widetilde E$ is related to the $4d$ one for states contributing to the index as 
\be
2\widetilde E=E-\hat r\,.
\ee The index is independent of $\beta$ and gets contributions only from states satisfying
\be
\widetilde E-R_H-R_C-j_2=0.
\ee

Let us comment on the fugacity $\t$.  Since it couples to the difference $J_t = J_3^H - J_3^C$, we see that, under mirror symmetry, where the two R-symmetries are exchanged, $\t$ will map to $\t^{-1}$.  This gives an indication as to whether a given duality is a mirror symmetry or not.

Let us mention useful examples of the ${\cal N}=4$  index. The  index of a free hyper-multiplet is given
 by\footnote{Here we should mention that in presence of magnetic charges $m$
 what is meant by $(-1)^F$ depends on the charges 
of the states, {\it i.e.} $F\to F+e\cdot m$ where $e$ is the ``electric'' charge of a  state~\cite{Dimofte:2011py}. This fact is crucial in obtaining correct indices transforming properly under dualities~\cite{Aharony:2013dha,Aharony:2013kma}. However, in the particular cases discussed
in this paper this  will not play any role: not to clatter notations the ``naive'' definition of $F$ is used 
which amounts to redefining the Cartan $U(1)$ fugacities $z\to (-1)^m z$.
}
\be
{\cal I}_{hyp}(z,m;\t,\q)=
\left(
\frac{\q^{\frac12}}{\t}
\right)^{\frac12|m|}\;
\prod_{\ell=0}^\infty\frac
{1-\t^{-\frac12}\q^{\frac34+\frac12|m|+\ell}\,z}
{1-\t^{\frac12}\q^{\frac14+\frac12|m|+\ell}\,z}
\frac
{1-\t^{-\frac12}\q^{\frac34+\frac12|m|+\ell}\,z^{-1}}
{1-\t^{\frac12}\q^{\frac14+\frac12|m|+\ell}\,z^{-1}}\,,
\ee
which can be written compactly using the $q$-Pochammer symbol, defined for $|q|<1$ by $(z;q)=\prod_{\ell=0}^\infty (1-z q^\ell)$ (we will often suppress the arguments $\t,\q$),
\be
{\cal I}_{hyp}(z,m)=
\left(
\frac{\q^{\frac12}}{\t}
\right)^{\frac12|m|}\;
\frac{
(\t^{-\frac12}\q^{\frac34+\frac12|m|} z^{\pm 1};\q)}
{(\t^{\frac12}\q^{\frac14+\frac12|m|}\,z^{\pm 1};\q)}\,.
\ee
Here and below, we use the convention that when a function appears with multiple choices of signs, the product is taken over all choices. The parameter $z$ is a fugacity for $U(1)_z$ symmetry under which the half-hypers have charges $\pm1$. The discrete parameter $m$ is the GNO charge of a background monopole configuration of $U(1)_z$.  These can be thought of as parameterizing the BPS configurations of a background vector multiplet which couples to the $U(1)$ global symmetry acting on the hyper.

When a $U(N)$ flavor symmetry is gauged in an $\mathcal{N}=4$ theory, the index is computed by the following matrix integral over holonomies $z_i$ in the unit circle, as well as a sum over integer GNO charges $m_i$,

\be
&&{\cal I}(a_j,n_j)=\sum_{\{m_i\}\in \Z} w^{\sum_i m_i}
\frac{1}{|W_{\{m_i\}}|}\;
\left(\frac{(\t\,\q^{\frac12};\q)}{(\t^{-1}\,\q^{\frac12};\q)}\right)^{N}\;
\left(
\frac{\q^{\frac12}}{\t}
\right)^{-\sum_{i<j}|m_i-m_j|}\;\times\\
&&\qquad\qquad\oint \prod_{i=1}^N
\frac{{z_i}^ndz_i}{2\pi i z_i} \prod_{i\neq j}(1-q^{\frac12|m_i-m_j|}z_i/z_j)
\frac{(\t\,\q^{\frac12+|m_i-m_j|/2}z_i/z_j;\q)}{(\t^{-1}\,\q^{\frac12+|m_i-m_j|/2}z_i/z_j;\q)}
{\cal I}_{matt}(z_i,\,m_i;a_j,n_j)\,.\nonumber
\ee
This includes the contribution of the $\mathcal{N}=2$ vector multiplet and the adjoint chiral multiplet.  Here ${\cal I}_{matt}(z_i,\,m_i;a_j,n_j)$ denotes the contribution of the matter hypermultiplets of the theory which may couple to the gauge field variables $\{z_i,m_i\}$ as well as flavor symmetry variables $\{a_j,n_j\}$.  In addition, the parameters $(w,n)$ represent the BPS configurations of a twisted vector multiplet which couples to the $U(1)_J$ symmetry. The discrete group $W_{\{m_i\}}$ is the subgroup of the Weyl group of $U(N)$ preserved
in the presence of gauge configurations with GNO charges $\{m_i\}$.  In the context where the index counts local operators in flat space, these GNO charges label monopole operators, which may further be dressed by the fields of the theory.

We should also comment that, in order to probe the index of the SCFT one obtains in the IR from such UV descriptions, as we would like to do, one must localize with respect to the correct superconformal R-symmetry \cite{Jafferis:2010un}.  Typically for $\mathcal{N}=4$ theories, the nonabelian structure of the R-symmetry group is sufficiently rigid that one can argue the same R-symmetry group acts in the UV and IR.  When this is the case, the partition function computations in this section will apply to the IR theory.  However, in \cite{Gaiotto:2008ak} it was shown that, for some theories, the so-call ``bad'' theories, this is not the case.  For such theories one in fact finds that the partition function computed as above, actually diverges.  In this paper we will restrict to theories which are ''good'' or ''ugly,'' in the sense of \cite{Gaiotto:2008ak}.  It is interesting to note that, although a theory may be bad, it may have a dual which is not bad, and so one can still probe its IR SCFT.  We will in fact see examples of this in what follows.

\subsection*{The ${\S^3}_b$ partition function}

The ${\S^3}_b$, or squashed sphere, partition function can also be obtained from the $4d$ index by  taking a limit where the radius of the $\S^1$ goes to zero~\cite{Dolan:2011rp,Gadde:2011ia,Imamura:2011uw} (see also~\cite{Aharony:2013dha}).  More precisely, we define $3d$ parameters $\gamma,\sigma$ in terms of the $4d$ parameters via:\footnote{The funny shift in the definition of $\gamma$ is analogous to the shift by $\q^{\frac{1}{2}}$ of in the definition of $\t$ in (\ref{indred}), and is made so that $\gamma$ transforms simply under mirror symmetry, namely, as $\gamma \rightarrow -\gamma$.}
\be
 t=e^{2\pi i  r_1 (\gamma + \frac{i}{2 r_3}(b+b^{-1}))},  \;\;\; z=e^{2\pi i  r_1 \sigma}, \;\;\; p=e^{2\pi b  r_1/r_3}, \;\;\; q=e^{2\pi b^{-1}   r_1/r_3} \,.
\ee
We then take the limit of the $4d$ index where $r_1$, which we identify with the radius of the $\S^1$, goes to zero.\footnote{We also should remove  certain divergent prefactors
which appear in the limit related to the gravitational anomaly, as in~\cite{Aharony:2013dha}.}  Here $b$ is the squashing parameter and  $r_3$ the radius of the ${\S^3}_b$.  The holonomies for global symmetries, {\it e.g.}, $z$, descend to $\mathcal{N}=2$  real mass parameter, $\sigma$, in the $3d$ limit.  In other words, we have picked a privileged $\mathcal{N}=2$ subalgebra and the real mass parameters correspond to VEVs of the scalars in the $\mathcal{N}=2$ background vector multiplet, while VEVs of other scalars in the $\mathcal{N}=4$ vector multiplet cannot be turned on without further breaking the supersymmetry.

Carrying out this procedure for a free hypermultiplet, we find it contributes the following factor to the ${\S^3}_b$ partition function:
\be
\label{s3hypdef}
{\cal Z}_{hyp}(\sigma;\gamma,b)=\Gamma_h(\frac12\,\omega + \frac12 \gamma \pm\sigma;\omega_1,\omega_2)\,.
\ee Here we have defined $\omega_1=ib, \omega_2=ib^{-1}$, and $\omega=\frac{1}{2} (\omega_1 + \omega_2)$, and $\Gamma_h(z;\omega_1,\omega_2)$ is the hyperbolic Gamma function, given by (for $\mbox{Im}(\omega_2/\omega_1)>0$):
\be
\label{ghdef}
\Gamma_h(z;\omega_1,\,\omega_2)=
e^{\frac{\pi i}{2\omega_1\omega_2}\left((z-\omega)^2-\frac{\omega_1^2+\omega_2^2}{12}\right)}\,
\frac{(e^{\frac{2\pi i}{\omega_1}(\omega_2-z)};e^{\frac{2\pi i\omega_2}{\omega_1}})}
{(e^{-\frac{2\pi i}{\omega_2}\,z};e^{-\frac{2\pi i\omega_1}{\omega_2}})}\,.
\ee

One gauges a $U(N)$ symmetry in an $\mathcal{N}=4$ theory by performing the following integral:
\be
{\mathcal Z}(m_a) =  \frac{1}{N!}\Gamma_h(\omega-\gamma)^{N} \int d^N \sigma e^{-2 \pi i \zeta \sum_i \sigma_i} \prod_{i \neq j} \frac{ \Gamma_h(\omega-\gamma+\sigma_i-\sigma_j;\omega_1,\omega_2)}{\Gamma_h(\sigma_i - \sigma_j;\omega_1,\omega_2)} Z_{matt}(\sigma;m_a)\,. \nonumber \\
\ee
where $\sigma=\{\sigma_i\}$, taking values in the Cartan of $U(N)$, is the background value of the scalar in the $\mathcal{N}=2$ gauge multiplet, and $m_a$ parameterizes the scalars in background multiplets coupled to flavor symmetries.  Also $\zeta$ parameterizes the scalar in a background twisted vector multiplet coupled to the $U(1)_J$ symmetry, {\it i.e.}, it is an FI parameter.  Here the numerator is the contribution of the adjoint chiral multiplet, and the denominator is that of the $\mathcal{N}=2$ gauge multiplet.  The latter can be simplified using:

\be \label{gaugsimp}
\prod_{i \neq j}  \frac{1}{\Gamma_h(\sigma_i - \sigma_j;\omega_1,\omega_2)} = \prod_{i<j} 2 \sin\frac{\pi}{\omega_1} (\sigma_i- \sigma_j)
2 \sin\frac{\pi}{\omega_2} (\sigma_j- \sigma_i) \nonumber\\\nonumber\\
=  \prod_{i<j} 2 \sinh \pi b (\sigma_i- \sigma_j)
2 \sinh \pi b^{-1} (\sigma_i- \sigma_j) \hspace{1cm} \ee

\

\subsection*{Holomorphic Blocks}

In \cite{Pasquetti:2011fj,Beem:2012mb} it was shown that, for a wide class of $\mathcal{N}=2$ three dimensional theories, both of the partition functions discussed above can be assembled from the same basic ingredient, the holomorphic block.  This can be thought of as a partition function on $D^2 \times \S^1$, with $D^2$ a two-dimensional disk.  Gluing two copies of this space along their boundary tori in two different ways, one recovers $\S^3_b$ and $\S^2 \times \S^1$, and, correspondingly, taking two kinds of ``fusions'' of two sets of holomorphic blocks, one can recover the two different partition functions.  Let us briefly review how this works, focusing on theories with $\mathcal{N}=4$ supersymmetry.

We will take the holomorphic block of a free hypermultiplet to be,
\be\label{hyperblocks}
 B_{hyp}(\z;\t,\q) = \frac{(\q^{\frac{3}{4}} \t^{-\frac{1}{2}} \z^{-1};\q)}{( \q^{\frac{1}{4}} \t^{\frac{1}{2}} \z^{-1};\q)}\,, \ee
where we have used the $q$-Pochammer symbol, defined for general $q$ by,
\be
 (z;q) = \left\{ \begin{array}{cc} \prod_{r=0}^\infty (1 -z q^r) &, \;\; |q|<1 \\ \prod_{r=0}^\infty (1 -z q^{-r-1})^{-1} &,\;\; |q|>1 \end{array} \right. 
\ee
To recover the partition functions, we take the product of two blocks with modularly transformed parameters.  For example, for the $\S^3_b$ partition function, one finds\footnote{ More precisely, one finds the partition function produced by the blocks come with an additional background off-diagonal Chern-Simons term, which we have included as a prefactor on the LHS.}  
\be
\label{s3hyperdef}
 e^{\frac{\pi i}{\omega_1 \omega_2} m(\gamma-\omega)} {\mathcal Z}_{hyp}(\sigma;\gamma,b) = B_{hyp}(\z;\t,\q) B_{hyp}(\tilde{\z};\tilde{\t},\tilde{\q}) \,,
\ee
where we define\footnote{Here we should specify that $\t = -e^{2\pi b \gamma}$ should be interpreted as $\t = e^{\pi i + 2\pi b \gamma}$, and similarly $\tilde{\t}=e^{\pi i + 2\pi b^{-1} \gamma}$, in order to fix branch cut ambiguities that will arise since $\t^{\frac{1}{2}}$ appears in many expressions.}
\be
\label{s3blockdef}
\z=e^{2 \pi b z},\;\;\; \t=-e^{2 \pi b \gamma}, \;\;\; \q=e^{2 \pi i b^2}\,,
\ee
and similarly for $\tilde{\z},\tilde{\t},\tilde{\q}$, with $b \leftrightarrow b^{-1}$.  One can check from (\ref{ghdef}) that this reproduces (\ref{s3hypdef}).  

\

\noindent Similarly, for the $\S^2 \times \S^1$ index, we have:
\footnote{Here, and throughout this paper,
 we do not consider turning on a flux for the symmetry with fugacity $\t$: this would be natural from the ${\cal N}=2$ perspective, since then $R_H-R_C$ is just a global flavor symmetry, but less natural
from the ${\cal N}=4$ perspective.}
\be
  (\q^{\frac{1}{4}} \t^{-\frac{1}{2}})^m {\mathcal I}_{hyp}(z,m;\t;\q) =  B_{hyp}(\z;\t,\q) B_{hyp}(\bar{\z};\bar{\t},\bar{\q})\,,\ee
where we now define:
\be
\label{s2s1blockdef}
\z = z \q^{\frac{m}{2}}, \;\;\; \bar{\z} = z^{-1} \q^{\frac{m}{2}}, \;\;\; \bar{\t} = \t^{-1} , \;\;\; \bar{\q} = \q^{-1}\,.
\ee
The block of the $\mathcal{N}=4$ adjoint chiral multiplet is given by:
\be B_V(\z,\t,\q)= \prod_\alpha (\q^{\frac{1}{2}} \t \z^\alpha;\q) \,.\ee

The free theories described above factorize into one pair of blocks, but, as was shown in~\cite{Beem:2012mb} and as
we will see concretely in an example below, the process of gauging a symmetry is more subtle than for the two partition functions described above, and for a gauge theory one finds the partition functions are written as a sum over blocks, {\it e.g.}, for a theory on $\S^2 \times \S^1$ with parameters $\{a_i,m_i\}$ for global symmetries and corresponding block parameters $\a_i$:
\be
\label{blockfact}
{\mathcal I}_{gauge\; theory}(\{a_i,m_i\})= \sum_{\alpha=1}^{\frak r} B_{\alpha}(\{ {\a}_i\}) B_{\alpha}(\{ \tilde{\a}_i \}) 
\ee
where $\frak r$ is roughly the number of (fully gapped) vacua of the theory at generic values of the mass parameters.  We will study this in detail  in what follows for the example of the $T[SU(2)]$ theory, {\it i.e.}, SQED with two flavors.

\

\subsection{Useful limits of the $3d$ partition functions}

There are several interesting limits of the $3d$ partition functions which one can discuss.
In these limits some of the parameters are sent to special values resulting in the
partition functions simplifying tremendously.

\

\subsection*{Limits of the index}

It is useful to define the following combinations of fugacities
\be x = \q^{\frac{1}{2}} \t,\;\;\;\; \tilde{x} = \q^{\frac{1}{2}} \t^{-1}\,.\ee
In terms of these fugacities the index~\eqref{index3d} can be written as
\be
{\cal I}(x,\tilde{x})=\Tr(-1)^F\, x^{\widetilde E-R_C}\;{\widetilde x}^{\widetilde E-R_H}\,  e^{-2\beta(\widetilde E-R_H-R_C-j_2)}\,.
\ee 
Under mirror symmetry $x$ and $\widetilde x$ are exchanged.
Note that 
\be\label{boundpos}
\widetilde E\geq R_{H,C}\,.
\ee This follows from unitarity: the eight supercharges of ${\cal N}=4$ supersymmetry anticommute
with their superconformal counterparts to,
\be
\widetilde E\pm R_H\pm R_C\pm j_2\geq0\, .
\ee 
The inequality~\eqref{boundpos}  makes it sensible to consider the limits of the index we are about to discuss.

\

\noindent We define Coulomb/Higgs limits as follows 
\be
Higgs\;:&\qquad \q\to 0,\t\to\infty\quad (\widetilde x\to 0),\qquad &\t\,\q^{\frac12}=x\;\;fixed\,,\\
Coulomb\;:&\qquad \q,\t\to 0\quad (x\to 0),\qquad &\frac{\q^{\frac12}}\t=\widetilde x\;\;fixed\,.\nonumber
\ee
The index we are computing is 
\be\label{CHlims}
{\cal I}^C(\tilde{x})=\Tr_{{\cal H}_C}\, (-1)^F\, {\widetilde x}^{\widetilde E-R_H}\,,\qquad 
{\cal I}^H(x)=\Tr_{{\cal H}_H}\, (-1)^F\, {x}^{\widetilde E-R_C}\,.
\ee Here ${\cal H}_{C,H}$ is the subspace of the Hilbert space on $\S^3$ with 
$\widetilde E=R_{C,H}$ respectively. The states which contribute to these limits of the index are annihilated by an additional supercharge. In the Coulomb limit this supercharge anticommutes 
with its hermitian conjugate to $\widetilde E+R_H-R_C+j_2$, and in the Higgs limit the extra supercharge anticommutes with the conjugate to $\widetilde E-R_H+R_C+j_2$.

 Note that since mirror symmetry exchanges R-symmetries acting on the Higgs and the Coulomb branches, $R_H$ with $R_C$, and so maps $\t \rightarrow \t^{-1}$, the Higgs (Coulomb)
limit of the index of a given theory maps to Coulomb (Higgs) limit of the mirror dual.
The index of a hypermultiplet becomes in these limits,
\be
&Higgs\;:& 
{\cal I}_{hyp}^H(z,m;x)=\delta_{m,0}\;
\frac
{1}
{1-x^{\frac12}\,z^{\pm1}}\,,\\
&Coulomb\;:& 
{\cal I}_{hyp}^C(z,m;\tilde{x})={\tilde x}^{\frac12|m|}\,.\nonumber
\ee
In particular, the former depends non-trivially only on the fugacity $z$, and the latter only on the flux $m$.  

When a symmetry is gauged we obtain in the Higgs limit (suppressing other flavor fugacities)
\be
&&{\cal I}^H(x)=
\frac{1}{N!}\;
(1-x)^{N}\;
\oint \prod_{i=1}^N
\frac{dz_i}{2\pi i z_i} \prod_{i\neq j}(1-z_i/z_j)
(1-x z_i/z_j)\;
{\cal I}_{matt}^H(\{z_i\};x)\,. \nonumber \\
\ee

\noindent In the Coulomb limit we get
\be
&&{\cal I}^C(\tilde{x})=\sum_{\{m_i\}\in \Z}
\frac{1}{|W_{\{m_i\}}|}\;
\left(\frac{1}{1-\tilde{x}}\right)^{N}\;
\tilde{x}^{-\sum_{i,j}|m_i-m_j|}\;\times\\
&&\qquad\qquad\oint \prod_{i=1}^N
\frac{dz_i}{2\pi i z_i} \prod_{i\neq j}
\left(\frac{1-z_i/z_j}{1- \tilde{x} z_i/z_j}\right)^{\delta_{m_i,m_j}}
{\cal I}^C_{matt}(\{m_i\};\tilde{x})\,.\nonumber
\ee 
Note that only the fugacity $w$ for the $U(1)_J$ symmetry appears in the Coulomb limit, and only the flux $n$ in the Higgs limit, which is the opposite behavior as for the flavor symmetry parameters.  This makes sense, since the former live in twisted vector multiplets and the latter in ordinary vector multiplets, and these two limits are exchanged by mirror symmetry.

In the Coulomb index the contribution
of the matter does not depend on the integration variables, and so the same integral appears for any choice of matter content.  We can evaluate this integral using\footnote{
Mathematically this is the Hall-Littlewood version of the Macdonald central term identity~\cite{Mac}.}
\be
\frac{\left(\frac{1}{1-\tilde{x}}\right)^{M}}{M!} \oint \prod_{i=1}^M \frac{dz_i}{2\pi i z_i} \prod_{i\neq j}
\left(\frac{1-z_i/z_j}{1-\tilde{x} z_i/z_j}\right)=\prod_{j=0}^{M-1}
\frac{1}{1-\tilde{x}^{j+1}}\,.
\ee 
From here one can immediately see that the Higgs/Coulomb limits give Hilbert series of the Higgs/Coulomb branch respectively  (see~\cite{Cremonesi:2013lqa,Cremonesi:2014kwa,Cremonesi:2014vla} for a recent
 discussion of the $3d$ Hilbert series of Higgs/Coulomb branch). We will discuss some examples in what follows.

\

\noindent Next we define two limits of the index which use the symmetries admitting certain relevant deformations,
\be
Mass_H\;:&&\qquad \t\to \q^{\frac12}\quad (\widetilde x=1)\,,\\
Mass_C\;:&&\qquad \t\to \q^{-\frac12}\,\quad (x=1)\, .
\ee 
  These two limits are again interchanged by mirror symmetry.
The index we are computing is 
\be
{\cal I}^{Mass_C}(\tilde{x})=\Tr\, (-1)^F\, {\widetilde x}^{\widetilde E-R_H}\,,\qquad 
{\cal I}^{Mass_H}(x)=\Tr\, (-1)^F\, {x}^{\widetilde E-R_C}\,.
\ee Note the traces here are over the whole Hilbert space on $\S^2$ unlike in~\eqref{CHlims}.
For the index of free hyper-multiplet these  limits give
\be
Mass_H\;:&&\qquad {\cal I}_{hyp}(z,m)=1,\\
Mass_C\;:&&\qquad {\cal I}_{hyp}(z,m)=\frac{{\tilde x}^{\frac12|m|}}{1-{\tilde x}^{\frac12|m|}z^{\pm1}}\,.
\ee The $Mass_H$ limit is consistent with giving all the hypermultiplets a complex mass.
Note that for the gauge theories these two   limits are a bit involved to compute since
the limit does not obviously commute with the infinite sum over the monopole sectors.

\

\noindent Finally there is the limit $\t=1$, or $x=\tilde{x}$, which is consistent with giving a mass to the adjoint chiral multiplet in the $\mathcal{N}=4$ vector multiplet.  This limit is taken to itself under mirror symmetry.

\

\subsection*{The partition function}

The only special limit of the $\S^3_b$ partition function we will discuss is the dimensional reduction
of the $4d$ Schur index, $q=t$ (or $p=t$)~\cite{Gadde:2011uv}.
This condition descends in the $\S^3_b$ partition to the limit $\gamma=\frac{i}{2}(b^{-1}-b)$ (or $\gamma=\frac{i}{2}(b-b^{-1})$ ).  We can see the simplification in this limit most easily at the level of the holomorphic blocks, namely, we recall from (\ref{s3hyperdef}):
\be
{\mathcal Z}_{hyp}(\sigma;\gamma,b) = e^{\pi i m(\gamma-\omega)}B_{hyp}(\z;\t,\q) B_{hyp}(\tilde{\z};\tilde{\t},\tilde{\q}) =  e^{\pi i m(\gamma-\omega)} \frac{(\q^{\frac{3}{4}} \t^{-\frac{1}{2}} \z^{-1};\q)}{(\q^{\frac{1}{4}} \t^{\frac{1}{2}} \z^{-1};\q)}\frac{(\tilde{\q}^{\frac{3}{4}} \tilde{\t}^{-\frac{1}{2}} \tilde{\z}^{-1};\tilde{\q})}{(\tilde{\q}^{\frac{1}{4}} \tilde{\t}^{\frac{1}{2}} \tilde{\z}^{-1};\tilde{\q})}\,.
\nonumber\\
\ee
In terms of the block variables the Schur limit is $\t=\q^{\frac{1}{2}} ,\tilde{\t} =\tilde{\q}^{-\frac{1}{2}}$, and one finds the blocks simplify as:
\be\frac{(\q^{\frac{3}{4}} \t^{-\frac{1}{2}} \z^{-1};\q)}{(\q^{\frac{1}{4}} \t^{\frac{1}{2}} \z^{-1};\q)} \rightarrow \frac{1}{1+z^{-1}}, \;\;\;\; \frac{(\tilde{\q}^{\frac{3}{4}} \tilde{\t}^{-\frac{1}{2}} \tilde{\z}^{-1};\tilde{\q})}{(\tilde{\q}^{\frac{1}{4}} \tilde{\t}^{\frac{1}{2}} \tilde{\z}^{-1};\tilde{\q})} \rightarrow 1\,.\ee
So that the partition function of the hyper becomes
\be
 {\mathcal Z}_{hyp}(\sigma;\gamma=\frac{i}{2}(b^{-1}-b),b) =  e^{-\pi m b} \frac{1}{1+z^{-1}} = \frac{1}{2 \cosh \pi b \sigma }\,.
\ee
In the other Schur limit (descending from $p=t$), the behavior of the two blocks is reversed, and one finds
\be {\mathcal Z}_{hyp}(\sigma;\gamma=\frac{i}{2}(b-b^{-1}),b) =   \frac{1}{2 \cosh \pi b^{-1} \sigma }\,.
\ee

We can similarly work out the contribution of the $\mathcal{N}=4$ vector multiplet in this limit, and one finds that, {\it e.g.}, for a $U(N)$ gauge theory:\footnote{Note the difference from (\ref{gaugsimp}) - this comes about because the adjoint chiral contributes as $\frac{b^N \sinh \pi b (\sigma_i-\sigma_j)}{\sinh \pi b^{-1} (\sigma_i-\sigma_j)}$ in this limit.}

\be
{\mathcal Z}(m_a;\gamma=\frac{i}{2}(b^{-1}-b)) =  \frac{b^N}{N!}\int d^N \sigma e^{-2 \pi i \zeta \sum_i \lambda_i}\prod_{i\neq j} 2 \sinh \pi b (\sigma_i - \sigma_j) {\mathcal Z}_{matt}(\sigma_j,m_a)
\ee
And similarly with $b \rightarrow b^{-1}$ for the other Schur limit.  Note that the dependence of the partition function on $b$ is trivial in this limit, as it can be absorbed into a rescaling of $\sigma$ and the real mass parameters by $b^{-1}$ and of the FI parameters by $b$.  This is analogous to a the fact that, in the $4d$ Schur limit, eg. $q=t$, the $\S^3 \times \S^1$ index becomes independent of $p$.

Note that mirror symmetry takes $\gamma \rightarrow - \gamma$, and so exchanges the two Schur limits.  This is compatible with the way we rescale mass and FI parameters in this limit.  

\

\subsection{Examples}

\subsubsection*{ {\bf (I)} - $U(1)$ SQED/ free hypermultiplet mirror symmetry}

Let us discuss some of these partition functions and limits in some simple examples, starting with $\mathcal{N}=4$ $U(1)$ SYM with one charge $1$ hypermultiplet, which we denote by $SQED_1$. This exhibits the most basic example of $\mathcal{N}=4$ mirror symmetry, being dual to a free (twisted) hypermultiplet.  The index of SQED is
\be
&& {\mathcal I}_{SQED_1}(w,n;\t,\q)=\frac{(\t\q^{\frac12};\q)}{(\t^{-1}\q^{\frac12};\q)}\,\sum_{\widetilde m\in {\Z}} \,
w^{\widetilde m} \,\left(\frac{\q^{\frac12}}{\t}\right)^{\frac{1}{2}|\widetilde m|} \oint\frac{dz}{2\pi i z}
z^{n}\, \frac{(\t^{-\frac12}\q^{\frac34+\frac12|\widetilde m|}z^{\pm1};\q)}
{(\t^{\frac12}\q^{\frac14+\frac12|\widetilde m|}z^{\pm1};\q)}\,.
\ee
The index of a free twisted hypermultiplet is
\be 
 {\mathcal I}_{hyp_t}(w,n;\t,\q)=\left(\q^{\frac12} \t \right)^{\frac{1}{2}|n|)} \, \frac{(\t^{\frac12}\q^{\frac34+\frac12|n|}w^{\pm1};\q)}
{(\t^{-\frac12}\q^{\frac14+\frac12|n|}w^{\pm1};\q)}\,.
\ee
Note that this differs from the index of an ordinary hypermultiplet by $\t \rightarrow \t^{-1}$.  These expressions can be shown to be equal \cite{Krattenthaler:2011da}.  We can see this explicitly in the Higgs/Coulomb limit.  For example, in the Higgs limit, we find:
\be
&& {\mathcal I}^H_{SQED_1}(w,n;x)=(1-x) \oint\frac{dz}{2\pi i z}\frac{z^n}{1-x^{\frac{1}{2}} z^{\pm 1}}  = x^{\frac{1}{2}|n|}\,,
\ee
which agrees with the Coulomb index of a free twisted hypermultiplet.  

\

\noindent We can also study the partition function of this theory.  This is given by:
\be {\mathcal Z}_{SQED_1}(\zeta;\gamma,b) =\Gamma_h(\omega-\gamma) \int d\sigma e^{-2 \pi i \zeta \sigma } \Gamma_h(\frac{\omega}{2} + \frac{\gamma}{2} \pm \sigma)\,. \ee
This is known \cite{FokkoThesis} to be equal to the partition function of a free twisted hypermultiplet
\be {\mathcal Z}_{hyp_t}(\zeta;\gamma,b) =\Gamma_h(\omega-\frac{\gamma}{2} \pm \zeta)\,.\ee
This can be seen explicitly in the Schur limit, $\gamma=ib$, where these become
\be &&  {\mathcal Z}_{SQED_1}(\zeta) \rightarrow \int d\sigma \frac{e^{-2 \pi i \zeta \sigma} }{2 \cosh \pi b \sigma}\,,\\
&&{\mathcal Z}_{hyp_t}(\zeta) \rightarrow \frac{1}{2 \cosh \pi b^{-1} \zeta}\,,\ee
and the equality of these follows from the fact $1/\cosh$ is fixed under the Fourier transform.  In fact, since we have seen above that the $b$-dependence can be removed by suitably rescaling parameters, the partition function in this limit essentially reduces to that of $\mathcal{N}=4$ theories on the round sphere, where many such checks have been performed in the literature.

\subsubsection*{{\bf (II)} - $\mathcal{N}=4$ Seiberg-like duality}

As an example of a non-mirror symmetry between $\mathcal{N}=4$ theories, we consider the duality discussed in \cite{Gaiotto:2008ak} between a $U(N)$ theory with $2N-1$ fundamental hypermultiplets (an ``ugly'' theory in their notation) and $U(N-1)$ with $2N-1$ fundamental hypermultiplets (a ``good'' theory) plus a decoupled free twisted hypermultiplet.  In the case $N=1$, this reduces to the duality above, and can be thought of as a mirror symmetry, but for general $N$, ordinary (as opposed to twisted) vector- and hypermultiplets appear in both the $U(N)$ and $U(N-1)$ gauge theories, so this is not a mirror symmetry.  We can see this explicitly by studying the index.

The index of a general $U(N_c)$ theory with $N_f$ fundamental hypermultiplets is given by:

\be {\mathcal I}_{U(N_c)_{N_f}}(w,n;\mu_a,\hat{m}_a) =
\left(\frac{(\t\,\q^{\frac12};\q)}{(\t^{-1}\,\q^{\frac12};\q)}\right)^{N_c}\;
\sum_{\{m_i\}}\frac{1}{|W_{\{m_i\}}|} w^{\sum_i m_i}\;\left(
\frac{\q^{\frac12}}{\t}
\right)^{-\sum_{i<j}|m_i-m_j|}\; \nonumber \\
 \oint \prod_{i=1}^{N_c}
\frac{{z_i}^n dz_i}{2\pi i z_i} \prod_{i\neq j}(1-q^{\frac12|m_i-m_j|}z_i/z_j)
\frac{(\t\,\q^{\frac12+|m_i-m_j|/2}z_i/z_j;\q)}{(\t^{-1}\,\q^{\frac12+|m_i-m_j|/2}z_i/z_j;\q)}
 \prod_{i=1}^{N_c} \prod_{a=1}^{N_f} I_{hyp}(z_i \mu_a,m_i+\hat{m}_a) \nonumber \\
\ee
Then the statement of the duality is that:\footnote{Note that we cannot see all the symmetries here, as there is an additional symmetry on the RHS theory which acts only on the free hyper which is realized as a hidden symmetry on the LHS theory.  What we observe here is just a particular codimension one slice in parameter space of the most general partition function with mass and FI deformations.}

\be {\mathcal I}_{U(N)_{2N-1}}(w,n;\mu_a,\hat{m}_a) = {\mathcal I}_{hyp_t}(w,n) {\mathcal I}_{U(N-1)_{2N-1}}(w,n;{\mu_a}^{-1},-\hat{m}_a) \ee
It is difficult to evaluate these integrals explicitly, but we have verified this in several examples by expanding both sides as a series to a high order in $\q$.  We can obtain analytic formulas in certain of the limits considered above.  For example, in the Higgs limit, the index of a $U(N_c)$ theory with $N_f$ flavors is:

\be \mathcal{I}^H_{U(N_c)_{N_f}}(n,\mu_a;x)=\frac{1}{N_c!} (1-x)^{N_c} \oint \prod_i \frac{dz_i}{2 \pi i z_i} {z_i}^n \frac{\prod_{i\neq j} (1 -z_i z_j^{-1})(1-x z_i {z_j}^{-1})}{\prod_{i=1}^{N_c} \prod_{a=1}^{N_f} (1-x^{\frac{1}{2}} (z_i \mu_a)^{\pm 1})} \ee
This can be computed by summing the finitely many poles that lie inside the unit circle.  Let us assume $n \geq 0$.  Then these lie at:

\be z_i = x^{\frac{1}{2}} \mu_{a(i)}^{-1} \ee
for some function $a(i):\{1,...,N_c\} \rightarrow \{1,...,N_f\}$.  When $N_c>N_f$, there are no such functions, and indeed in these cases there are not enough poles to soak up the contour integrals, and the index vanishes, reflecting supersymmetry breaking in these theories.  Thus let us assume $N_c \leq N_f$.   Then we should take the residues at these poles and sum over all choices of the function $a(i)$.  For example, suppose $a(i)=i$.  Then, after some cancellations between the numerator and the denominator, the residue is computed to be:

\be \mbox{Res}_{ z_j \rightarrow x^{\frac{1}{2}} \mu_{j}^{-1}} \mathcal{I}^H_{U(N_c)_{N_f}}(n,\mu_a;x) = \frac{1}{N_c!} \frac{ x^{nN_c /2} \prod_i {\mu_i}^{-n}}{\prod_{i=1}^{N_c}
 \prod_{a=N_c+1}^{N_f} (1-\mu_i {\mu_a}^{-1}) (1-x \mu_a {\mu_i}^{-1}) } \ee
It remains to sum over the choices of $a(i)$.  We can break this up into two parts: the choices of image $a(\{1,...,N_c\})$, of which there are $\binom{N_f}{N_c}$, and a choice of permutations $\pi \in S_{N_c}$, which takes $a(i) \rightarrow a(\pi(i))$ while preserving the image.  Note the expression above is invariant under such a permutation, so the sum over these simply eliminates the factor of $\frac{1}{N_c!}$.  Thus we find:

\be\mathcal{I}^H_{U(N_c)_{N_f}}(n,\mu_a;x) = \sum_{A \subset \{1,...,N_f\}, |A|=N_c} 
 \frac{ x^{nN_c /2} \prod_{a \in A} {\mu_a}^{-n}}{\prod_{a \in A,b\in A^c} (1-\mu_a {\mu_b}^{-1}) (1-x \mu_b {\mu_a}^{-1}) } \ee

Now consider the expression we get after taking $N_c \rightarrow N_f-N_c$.  The sum is over an isomorphic set, after exchanging the roles of $A$ and $A^c$, and we find:

\be \mathcal{I}^H_{U(N_f-N_c)_{N_f}}(n,\mu_a;x)= \sum_{|A^c|=N_c} 
 \frac{ x^{n(N_f-N_c) /2} \prod_{a=1}^{N_f} {\mu_a}^n \prod_{a \in A^c} {\mu_a}^{-n}}{\prod_{a \in A^c,b\in A} (1-\mu_a {\mu_b}^{-1}) (1-x \mu_b {\mu_a}^{-1}) } \nonumber \ee

\be =  \bigg(\prod_{a=1}^{N_f} {\mu_a}^n \bigg) x^{n(N_f-2 N_c)/2} \mathcal{I}^H_{U(N_c)_{N_f}}(n,{\mu_a}^{-1};x) \ee
The prefactor can be thought of as a contact term, and becomes $1$ if we impose that the fugacities are valued in $SU(N_f)$, as we will do from now on.  Also, recall we had assumed $n \geq 0$; for $n<0$, we can simply start by performing a change of variables $z_i \rightarrow {z_i}^{-1}$, and will obtain an expression of the above form, but with $n \rightarrow -n$.  Thus the general relation is:

\be \label{sld}
\mathcal{I}^H_{U(N_f-N_c)_{N_f}}(n,\mu_a;x) = x^{|n|(N_f-2 N_c)/2} \mathcal{I}^H_{U(N_c)_{N_f}}(n,{\mu_a}^{-1};x) \ee
In particular, in the case $N_f=2N_c-1$, we find:

\be \mathcal{I}^H_{U(N)_{2N-1}}(n,\mu_a;x) =  x^{|n|/2} \mathcal{I}^H_{U(N-1)_{2N-1}}(n,{\mu_a}^{-1};x) \ee
which is precisely the duality between the good and ugly theory, where we recognize the $x^{|n|/2}$ factor as the contribution of a free twisted hyper in the Higgs limit (equivalently, of an ordinary hyper in the Coulomb limit).  The equality of the indices of these theories in the Coulomb limit was checked in \cite{Cremonesi:2013lqa}.  The result (\ref{sld}) is also suggestive of a more general $3d$ $\mathcal{N}=4$ Seiberg-like duality between good and bad theories proposed in \cite{Gaiotto:2013bwa,Yaakov:2013fza}

One can similarly check that the $\S^3_b$ partition functions of these theories are equal.  In certain subsets of parameter space these identities reduce to similar ones which have been proven for $\mathcal{N}=2$ Aharony duality \cite{Aharony:1997gp} in \cite{FokkoThesis,Willett:2011gp}, but for generic parameters they are distinct.

\

\section{Down the rabbit hole}

In this section we will study a class of $3d$ $\mathcal{N}=4$ theories which come from dimensional reduction of the $\mathcal{N}=2$ class ${\cal S}$ theories in four dimensions.  We will refer to these three dimensional theories as theories of class ${\frak s}$.  Specifically, we will focus for simplicity on the $A_1$ class of theories, in which case these are $SU(2)$ quiver gauge theories, although much of what we say will generalize to the case of higher rank.  In addition to these theories, an important role will be played by the so-called $T[SU(2)]$ theory, or $\mathcal{N}=4$ SQED with two hypermultiplets.
This theory  appears in several interesting contexts. For example it is a building block in the construction of the mirror duals of the class ${\frak s}$ theories, the so-called star-shaped quiver theories. It is also the theory living on the S-duality domain wall  relating two copies of ${\cal N}=4$
theory in $4d$~\cite{Gaiotto:2008ak}. In the latter context the $\S^3$ partition function
of this theory plays a role of the duality kernel in $\S^4$ partition function computations~\cite{Hosomichi:2010vh}.

\

\subsection{Partition Functions of $T(SU(2))$}

Let us start with the $T[SU(2)]$ model. It is a $U(1)$ gauge theory with two charged ${\cal N}=4$ hypermultiplets.  The index of this theory, which we write in a suggestive notation whose purpose will become clear below, is given by (see {\it e.g.} \cite{Gang:2013sqa})
\be\label{TSU2p}
&&\psi_{\q,\t}(a,m|b,n)=\frac{(\t\q^{\frac12};\q)}{(\t^{-1}\q^{\frac12};\q)}\,\times\\
&&\sum_{\widetilde m\in {\Z}+\epsilon(m)} \,
b^{2\widetilde m} \,\left(\frac{\q^{\frac12}}{\t}\right)^{\frac{1}{2}(|\widetilde m + m|+|\widetilde m - m|)} \oint\frac{dz}{2\pi i z}
z^{2n}\, \frac{(\t^{-\frac12}\q^{\frac34+\frac12|\pm \widetilde m\pm m|}z^{\pm1}a^{\pm 1};\q)}
{(\t^{\frac12}\q^{\frac14+\frac12|\pm \widetilde m\pm m|}z^{\pm1}a^{\pm1};\q)}\,,\nonumber
\ee where
\be
\epsilon(m)=\frac{1-(-1)^{2m}}{4}\,.
\ee
Here, the fugacity $a$ labels the $su(2)_V$ symmetry acting on the Higgs branch; more precisely it corresponds to the $u(1)$ Cartan, which is normalized such that the quarks, which live in the fundamental representation, have charges $\pm 1$.  In addition, $b$ labels the $u(1)_J$ topological symmetry acting on the Coulomb branch, which is the Cartan of an enhanced $su(2)_J$ symmetry which appears in the IR. Finally, the integers $n$ and $m$ are GNO monopole charges for these two $su(2)$s.  For monopole configurations which are well-defined for $SU(2)$, these must be integers, while for $SO(3)$ monopoles one may also allow half-integers.  Naively one must allow only integer fluxes for the $su(2)_V$ fugacity, since the quarks sit in the fundamental representation.  However, one may also allow half-integer fluxes if one simultaneously takes the flux of the gauged $U(1)$ to be in $\mathbb{Z}+\frac{1}{2}$.  This choice is implemented above in the definition of $\epsilon(m)$.  In other words, while the flavor symmetry is naively $SU(2)$, since the $\mathbb{Z}_2$ center corresponds to a gauge symmetry, one can in fact consider it to be $SO(3)$, and correspondingly couple it to background $SO(3)$ gauge field configurations.  One can similarly take $SO(3)$ fluxes for the $su(2)_J$ symmetry.

\

\noindent The theory $T[SU(2)]$ has a mirror dual which is the same theory except with twisted, rather than ordinary, vector and hypermultiplets. For the index this implies,

\be\label{mirrorTSU}
\psi_{\q,\t}(a,m|b,n)=
\psi_{\q,\t^{-1}}(b,n|a,m)\,.
\ee Let us check this equality in the Higgs/Coulomb limit. In the Higgs limit of the left-hand side we get

\be
\psi^H_{x}(a,m|b,n)&=&(1-x)\;\delta_{m,0}\;\oint\frac{dz}{2\pi i z}z^{2n}\frac{1}{1-x^{\frac12}a^{\pm1}z^{\pm1}}=\\
&=&\delta_{m,0}\frac{x^{|n|} \left(a^{-2n-1}-a^{2 n+1} +x a^{2n-1}-x a^{1-2n} \right)}
{\left(a^{-1}-a\right)
 (1-x a^{-2}) \left(1-x a^2 \right)}\,.\nonumber
\ee For the right-hand side we compute the Coulomb index

\be
\psi^C_{\tilde{x}}(b,n|a,m)&=&\frac1{1-\tilde{x}}\sum_{\widetilde m\in {\Z}+\epsilon(n)} \,
a^{2\widetilde m} \,\tilde{x}^{\frac{1}{2}|(|n+\widetilde m|+|n-\widetilde m|)} \oint\frac{dz}{2\pi i z}
z^{2m}=\\
&=&\delta_{m,0}\frac{\tilde{x}^{|n|} \left(a^{-2n-1}-a^{2 n+1} +\tilde{x} a^{2n-1}- \tilde{x} a^{1-2n} \right)}
{\left(a^{-1}-a\right)
 (1-\tilde{x} a^{-2}) \left(1-\tilde{x} a^2 \right)}\,.\nonumber
\ee
We see explicitly that \eqref{mirrorTSU} is satisfied. We also note that 

\be\label{legHL}
&&\psi^H_{x}(a,m|b,n)= \delta_{m,0}\,x^{|n|}\;
\frac{1}{1-x\,a^{\pm2}}\; \chi^{(HL)}_{|2n|}(a;x)\,,\\
&&\chi^{(HL)}_{2n}(a;x)=
\chi_{2n}(a)-x\,\chi_{2n-2}(a)\,.\nonumber
\ee Here $\chi_{n}(a)$ are $A_1$ Schur, and $\chi^{(HL)}_{2}(a;x)$ $A_1$ Hall-Littlewood, polynomials.  This is equivalent to the observation in \cite{Cremonesi:2014kwa} that the Coulomb branch Hilbert series of the $T[SU(2)]$ theories, and other $T[G]$ theories, are given by Hall-Littlewood polynomials.

The theory $T[SU(2)]$ can be obtained as a theory living on the domain wall between two
 $su(2)$ ${\cal N}=4$ SYM theories in $4d$ related by S-duality~\cite{Gaiotto:2008ak}. The $su(2)_J\times su(2)_V$ flavor symmetry couples to the bulk gauge symmetries on either side of the wall.  Taking the global structure of the gauge group into account, we recall \cite{Aharony:2013hda} that, for ${\mathcal N}=4$ SYM, the $SU(2)$ theory is mapped under S-duality to the $SO(3)_+$ theory ({\it i.e.}, the $SO(3)$ theory where we include the basic 't Hooft loop operator), and the $SO(3)_-$ theory ({\it i.e.}, the $SO(3)$ theory where we include the basic dyonic 't Hooft-Wilson loop operator), is mapped to itself.  We can see this structure at the level of the index of the $T[SU(2)]$ theory.  Namely, we can view the states of $T[SU(2)]$ as operators living at the ends of $4d$ line operators~\cite{Dimofte:2011py} by performing a Fourier transform of the index with respect to the two flavor fugacities $a$ and $b$ to an electric charge basis ($e_a$ and $e_b$ respectively):
\be
\widetilde \psi_{\q,\t}(e_a,m|e_b,n)=\oint\frac{db}{2\pi i b}b^{-2e_b}
\oint\frac{da}{2\pi i a}a^{-2e_a}\psi_{\q,\t}(a,m|b,n)\,.
\ee The Fourier transform with respect to $b$ kills the sum over $\widetilde m$ in~\eqref{TSU2p}
and is non zero only if $\epsilon(m)=\epsilon(e_b)$. Going to the mirror frame we analogously deduce
that the Fourier transform with respect to $a$ is non zero only when $\epsilon(e_a)=\epsilon(n)$. Thus
we can write that,
\be
\widetilde\psi_{\q,\t}(e_a,m|e_b,n)\propto \delta_{\epsilon(m),\epsilon(e_b)}\;\delta_{\epsilon(e_a),\epsilon(n)}\,,
\ee  
In other words, states with odd electric charge $e_a$ and integer magnetic charge $m$ have even electric charge $e_b$ and half-integer magnetic charge $n$, {\it i.e.}, Wilson line operators map to 't Hooft line operators in $4d$.  States with $e_a$ odd and $m$ half-integer have also $e_b$ odd and $n$ half-integer, {\it i.e.}, 't Hooft-Wilson
line operators map to themselves.  This correlation between the charges of the two flavor groups was also discussed in~\cite{Gang:2012ff}.  

\

\noindent We can also consider the $\S^3_b$ partition function of $T[SU(2)]$, which is given by:
\be \phi_{\gamma,b}(m|\zeta) = \Gamma_h(\omega - \gamma) \int d\sigma e^{-4 \pi i \zeta \sigma} \Gamma_h(\frac{\omega}{2} + \frac{\gamma}{2} \pm \sigma \pm m)\,.\ee
This simplifies in the Schur limit, $\gamma=\frac{i}{2}(b-b^{-1})$, to
\be&& \phi_{\gamma,b}(m|\zeta) \to  \int d\sigma \frac{e^{-4 \pi i \zeta \sigma}}{2 \cosh \pi b (\sigma \pm m)}\\
&&\qquad\qquad = \frac{\sin 4 \pi \zeta m}{2 \sinh 2 \pi b^{-1} \zeta \; \sinh 2\pi b m } \,.\nonumber\ee
In this form we explicitly see the mirror symmetry $m \leftrightarrow \zeta$ and $\gamma\rightarrow -\gamma$ (which in this limit become $b \rightarrow b^{-1}$).

Let us note an important property of the $T[SU(2)]$ theories related to gluing two such theories together by gauging the diagonal sum of one of the $SU(2)$ flavor symmetries of each.  We claim that such an operation produces a ``delta functional'', which sets the two remaining $SU(2)$ flavor symmetries to be equal.  This can be seen explicitly at the level of the partition function, for example, in the Schur limit, gauging two $T[SU(2)]$ theories together is accomplished by:

$$ \frac{1}{2} \int dm \frac{\sin 4 \pi \zeta_1 m}{2 \sinh 2 \pi b^{-1} \zeta_1 \; \sinh 2\pi b m } (2 \sinh 2 \pi b m)^2 \frac{\sin 4 \pi \zeta_2 m}{2 \sinh 2 \pi b^{-1} \zeta_2 \; \sinh 2\pi b m } $$

\be\label{ortho} = \frac{1}{2} \int dm \frac{\sum_{\epsilon_1,\epsilon_2 \in \{\pm1\}}\epsilon_1 \epsilon_2 e^{4 \pi i m(\epsilon_1 \zeta_1 + \epsilon_2 \zeta_2)} }{\sinh 2 \pi b^{-1} \zeta_1 \;\sinh 2 \pi b^{-1} \zeta_1}  = \frac{1}{2}\sum_{\pm} \frac{\delta(\zeta_1 \pm \zeta_2)}{(2 \sinh \pi b^{-1} \zeta_1)^2} \ee
We recognize the factor multiplying the delta function as the inverse of the contribution of the vector multiplet in the other Schur limit, and we claim that in general, for both the index and partition function, we find a delta function times the inverse of the contribution of a twisted $SU(2)$ vector multiplet. 
We can think of this as an orthogonality property for the $T[SU(2)]$ partition functions. This property was noticed in \cite{Benvenuti:2011ga,Nishioka:2011dq}, where they also found analogous expressions for the higher $T[SU(N)]$ theories. A physical interpretation is that, although, a priori, the theory one obtains by gluing two $T[SU(2)]$ theories has an $su(2)\times su(2)$ symmetry, only the diagonal 
combination of the two $su(2)$s is a good symmetry  in the IR.\footnote{See~\cite{Spiridonov:2014cxa}
 for a similar effect at the level of the index in $4d$.} 
 We will see why this property is important for the $3d$ class ${\frak s}$ theories and their mirrors below.

\

\subsection*{Blocks of T[SU(2)]}

Both the $\S^2 \times \S^1$ index and $\S^3_b$ partition function of the $T[SU(2)]$ can be expressed in terms of the same holomorphic blocks, although in a non-unique way.  To see how this works, let us first define
\be
\label{tsu2blocks}
C^{T[SU(2)]}_{\pm}(\a,\b,\t,\q) = \sum_{j=0}^\infty (\b^2 \q^{\frac{1}{2}} \t^{-1})^j \frac{(\t \q^{\frac{1}{2}} ;\q) (\q^{1+j};\q) (\q^{1+j} \a^{\pm 2};\q)}{(\t \q^{\frac{1}{2}+j};\q)(\t \q^{\frac{1}{2}+j} \a^{\pm 2};\q)}\,.
\ee
Then we recover the index of $T[SU(2)]$ by (using the notation of (\ref{s2s1blockdef})):
\be
\psi_{\q,\t}(a,m|b,n) = \sum_\pm (\q^{\frac{1}{2}} \t^{-1} b^2)^{m/2} (\q^{\frac{1}{2}} \t a^2)^{n/2}  C^{T[SU(2)]}_{\pm}(\a,\b,\t,\q) C^{T[SU(2)]}_{\pm}(\bar{\a},\bar{\b},\bar{\t},\bar{\q}) \,.
\ee This expression can be directly obtained by evaluating the contour integral in~\eqref{TSU2p} by computing
residues of the integrand.
We similarly recover the $\S^3_b$ partition function via
\be
\phi_{\gamma,b}(m|\zeta) = \sum_\pm e^{-\pi i (2 \zeta -\gamma+\omega)(\pm 2 m + \gamma+\omega)}  C^{T[SU(2)]}_{\pm}(\a,\b,\t,\q) C^{T[SU(2)]}_{\pm}(\tilde{\a},\tilde{\b},\tilde{\t},\tilde{\q})\,.
\ee
This is not a complete factorization of these partition functions, because of the extra factors which appear multiplying the $C^{T[SU(2)]}_{\pm}$.  These factors look heuristically as contributions of Chern-Simons terms in the flavor symmetries, and can also be factorized using the theta function
\be \hat \theta(z;q) =(-q^{\frac{1}{2}} z;q)(-q^{\frac{1}{2}}z^{-1};q) \,,\ee
and the basic fusion relation:
\be
\label{thetafusion}
 \S^3_b: \; \hat\theta(\frak z;\frak q) \hat\theta(\tilde{\frak z};\tilde{\frak q})
 = e^{\pi i \sigma^2}, \qquad\qquad
\S^2 \times \S^1: \; \hat\theta(\frak z;\frak q) \hat\theta(\bar{\frak z};\bar{\frak q}) = z^{-m} \,.
\ee
Thus we can recover both factors by defining blocks:
\be
\label{bwt}
B^{T[SU(2)]}_{\pm}(\a,\b,\t,\q)  = \frac{\hat\theta(-\q^{\frac{1}{4}} \b \t^{-\frac{1}{2}};\q) \hat\theta(\q^{\frac{1}{2}} \a^2 \t;\q)}{\hat\theta(-\q^{\frac{3}{4}} \a^2 \b \t^{\frac{1}{2}};\q)}  C^{T[SU(2)]}_{\pm}(\a,\b,\t,\q)  \,.
\ee
However, this choice of theta functions is not unique, {\it e.g.}, one can consider a similar choice with $\a \leftrightarrow \b$ and $\t \leftrightarrow \t^{-1}$.  We should also emphasize that, since the fusion relation (\ref{thetafusion}) only holds for the index for integer $m$, with the choice of theta functions in (\ref{bwt}) we only recover the index of $T[SU(2)]$ for integer $n$ (but any half-integer $m$), and similarly only for integer $m$ with the choice with $\a \leftrightarrow \b$ and $\t \leftrightarrow \t^{-1}$.  

We can consider different limits at the level of the blocks. However, some of the limits are incompatible with different gluings. Another way to put this is that the limits correlate between the 
different block variables. For example, taking Higgs/Coulomb limit we take $\q\to 0$,
and thus also ${\frak a},\,\bar{\frak a}\to0$ but keep ${\frak a}/\bar{\frak a}$ fixed.
Moreover, ${\frak a}\bar{\frak a} \q^{-m}=1$.

\
\

\subsection{Class ${\frak s}$ theories and their star-shaped quiver mirrors}

Let us now consider the dimensional reduction of theories of class ${\cal S}$ and their mirror duals, the star-shaped quiver theories, which were described in \cite{Benini:2010uu}.  We start with building block of the $A_1$ theories of class ${\frak s}$, the $T_2$ theory.  The $T_2$ model is 
a free bi-fundamental $SU(2)\times SU(2)$ hypermultiplet.  The index of the $T_2$ theory is given by

\be
{\cal I}^{T_2}(\{[b_i,n_i]\}_{i=1}^3;\t,\q)=
\left(\frac{\q^{\frac12}}{\t}\right)^{\frac14\sum_{s_i=\pm1}|\sum_{\ell=1}^3s_\ell n_\ell|}
\prod_{s_i=\pm1}
\frac{(\t^{-\frac12}\q^{\frac{3}{4}+\frac12|\sum_{i=1}^3s_i n_i|}\prod_{i=1}^3b_i^{s_i};\q)}
{(\t^{\frac12}\q^{\frac14+\frac12|\sum_{i=1}^3s_i n_i|}\prod_{i=1}^3b_i^{s_i};\q)}\,.\nonumber\\
\ee  
The mirror dual of this theory is the star-shaped quiver built from three copies of $T[SU(2)]$ theory with 
the diagonal $su(2)_v$ flavor symmetry acting on the Higgs branches gauged.
The gauge group can be either $SU(2)$ or $SO(3)$.
The index of the star-shaped quiver with $SU(2)$ gauged in the central node is given by

\be
&&{\cal I}^{SSQ}(\{[b_i,n_i]\}_{i=1}^3;\t,\q)=\\
&&\; \frac{1}{2} \sum_{\widetilde m\in\Z} \,
\left(\frac{\q^{\frac12}}{\t}\right)^{-2|\widetilde m|}\, \oint\frac{dz}{2\pi i z}
(1-\q^{|\widetilde m|}z^{\pm2})\prod_{\ell=1}^3
\psi_{\q,\t}(b_\ell,n_\ell|z,\widetilde m)
\frac{(\t^{}\q^{\frac12+|\widetilde m|}z^{\pm2};\q)}
{(\t^{-1}\q^{\frac12+|\widetilde m|}z^{\pm2};\q)}
\frac{(\t^{}\q^{\frac12};\q)}
{(\t^{-1}\q^{\frac12};\q)}\,.\nonumber
\ee
One can also consider the index of the star-shaped quiver with $SO(3)$ gauged in the central node is given by

\be
&&\widetilde{\cal I}^{SSQ}(\{[b_i,n_i]\}_{i=1}^3;\t,\q)=\\
&&\; \frac{1}{2} \sum_{\widetilde m\in\frac{1}{2}\Z} \,
\left(\frac{\q^{\frac12}}{\t}\right)^{-2|\widetilde m|}\, \oint\frac{dz}{2\pi i z}
(1-\q^{|\widetilde m|}z^{\pm2})\prod_{\ell=1}^3
\psi_{\q,\t}(b_\ell,n_\ell|z,\widetilde m)
\frac{(\t^{}\q^{\frac12+|\widetilde m|}z^{\pm2};\q)}
{(\t^{-1}\q^{\frac12+|\widetilde m|}z^{\pm2};\q)}
\frac{(\t^{}\q^{\frac12};\q)}
{(\t^{-1}\q^{\frac12};\q)}\,.\nonumber
\ee
Note the only difference is in which monopole configurations we allow: only integer fluxes for $SU(2)$, and also half-integers for $SO(3)$.  

We claim in fact the correct dual of this theory is the star-shaped quiver with the central node gauged as an $SO(3)$ symmetry.  Namely, one can verify that the following equality holds

\be
&&{\cal I}^{T_2}(\{[b_i,n_i]\}_{i=1}^3;\t,\q)=
\widetilde {\cal I}^{SSQ}(\{[b_i,n_i]\}_{i=1}^3;\t^{-1},\q)\,, \nonumber \\
\ee 
Note that, in the $T_2$ theory, since the matter is in the trifundamental representation, one can consider the flavor symmetry to be $SU(2)^3/H$, where $H$ is the $\mathbb{Z}_2^2$ subgroup which acts as the center on an even number of the $SU(2)$ factors.  This means we can take the any fluxes $n_i$ such that the sum $\sum_{i=1}^3 n_i$ is even. 

For the star-shaped quiver with the central node gauged as an $SU(2)$ symmetry, one finds the following equality:
\be
\frac{1}{2}\left({\cal I}^{T_2}(\{[b_i,n_i]\}_{i=1}^3;\t,\q)+
{\cal I}^{T_2}(\{[-b_i,n_i]\}_{i=1}^3;\t,\q)\right)=
{\cal I}^{SSQ}(\{[b_i,n_i]\}_{i=1}^3;\t^{-1},\q)\,.
\ee 
This corresponds to a duality between this star-shaped quiver theory and the $T_2$ theory with the $\Z_2$ center gauged. \footnote{In other words, we are passing from an $SU(2)$ gauge theory to a $SU(2)/\mathbb{Z}_2$ gauge theory by gauging a discrete $\mathbb{Z}_2$ global symmetry, as in \cite{Aharony:2013kma,Kapustin:2014gua}.}

When computing the $\S^3$ partition function one can see that the partition function of the of the $T_2$ theory differs
from the partition function of the star-shaped quiver with $SU(2)$ gauged by a factor of two~\cite{Benvenuti:2011ga}, owing to this $\mathbb{Z}_2$ gauging, but agrees with the star-shaped quiver with $SO(3)$ gauged.

This mirror duality generalizes to all theories of class ${\cal S}$.
The mirror dual of class ${\cal S}$ theory corresponding to genus $g$ surface with $s$ punctures is a star-shaped quiver with $s$ copies of the $T[SU(2)]$ theory glued together by gauging
an $su(2)$ global symmetry with $g$ hypermultiplets in the adjoint representation of the gauge group
added in. The index of the star-shaped quiver has the following form
\be\label{mirroreq}
&&{\cal I}_{g,s}(\q,\t^{-1})=\\
&&\qquad\qquad\sum_{m=-\infty}^\infty \oint\frac{dz}{4\pi i z}
(1-\q^{|m|}z^{\pm2})\, I_{V,adj}(z,m) \left(I_{H,adj}(z,m)\right)^{g}\prod_{i=1}^s
\psi_{\q,\t}(b_i,n_i|z,m)\,.\nonumber\ee

The striking structural resemblance of (\ref{mirroreq}) to the $4d$ partition function~\eqref{maineq} is not a coincidence, as we will now discuss.

\subsection{Line operators, difference operators, and eigenfunctions}

In $4d$ the building blocks of the partition function computation are eigenfunctions, $\hat\psi_\lambda(a)$,
of certain difference operators as we reviewed in section~\ref{sec4d}. The structural similarity
of \eqref{maineq} and \eqref{mirroreq} suggests that we should identify after the dimensional
 reduction~\cite{Nishioka:2011dq} (see also~\cite{Gaiotto:2012xa,Bullimore:2014nla}),
\be
\hat\psi_\lambda(a)  \qquad\rightarrow \qquad \psi(a,m|b,n)\quad (\text{or} \;\; \phi(a|b))\,.
\ee
The labels of the eigenfunctions $\lambda$ become parameters $\{b,n\}$ or $b$ depending on whether 
we are computing the $\S^2\times \S^1$ or the $\S^3$ partition function in $3d$. We might expect 
that the partition functions of the $T[SU(2)]$ theory are eigenfunctions of the dimensional reductions
of the $4d$ difference operators. Such operators introduce surface defects in $4d$ and their dimensional reduction introduces line defects in $3d$. We now proceed to discuss how this comes 
about.

\

\subsection*{The $\S^3_b$ partition function}

The difference operators introducing the surface defects in $4d$ reduce in $3d$ to operators introducing line defects.  Let us first consider the reduction to the $\S^3_b$ partition function.  The basic difference operator on $\S^3 \times \S^1$ introduces a surface defect on one of the equators of $\S^3$,
 $z_1=0$ or $z_2=0$, and wrapping the temporal $\S^1$ direction. We reduce by shrinking the temporal circle. Thus in $3d$ we obtain a difference operator which introduces a single line
defect wrapping one of the equators of $\S^3_b$. A general difference operator
discussed in~\cite{Gaiotto:2012xa}  introduces defects labeled by symmetric representations on both equators.\footnote{See~\cite{Bullimore:2014nla} for a generalization to other representations for the higher rank cases.}  

After dimensional reduction the two basic difference operators~\eqref{S3Op1} act 
act on a real mass parameter $m$ in the $\S^3_b$ partition function, and are given by \cite{Gaiotto:2012xa}
\be
\label{s3difdef}
 \mathcal{O}^{m}_{(0,1)} = \frac{\sinh \pi b ( \frac{i (b-b^{-1})}{2}-\gamma + 2m)}{\sinh 2 \pi b m} \Delta_{m \rightarrow m + \frac{i b}{2}} - \frac{\sinh \pi b ( \frac{i (b-b^{-1})}{2}-\gamma - 2m)}{\sinh 2 \pi b m}\Delta_{m \rightarrow m - \frac{i b}{2}} \nonumber \\
\ee
as well as $\mathcal{O}^{m}_{(1,0)}$, which one obtains by taking $b \leftrightarrow b^{-1}$ in this expression.  Note that this operator is known as the Macdonald operator in mathematics literature.  When the three dimensional theory in question arises as a boundary of a four dimensional theory, such operators appear when one collides an 't Hooft loop with the boundary \cite{Dimofte:2011py}.  

\

\subsection*{The $\S^2 \times \S^1$ index}

Next we reduce to the $\S^2 \times \S^1$ index by considering the $r \rightarrow \infty$ limit of the lens index ({\it i.e.}, the $\S^3/\Z_r \times \S^1$  partition function).  Difference operators introducing surface defects in the lens index were studied in~\cite{Razamat:2013jxa}.  Here one does not have the basic operators $\mathcal{O}_{(1,0)}$ and  $\mathcal{O}_{(0,1)}$, but instead only their product survives the orbifold projection.  Thus the basic difference operator we obtain by dimensional reduction actually introduces a pair of line defects in $3d$.  The surface defects become line defects spanning $\S^1$ and sitting at 
the two poles of $\S^2$.  The explicit difference operator~\eqref{LensOp1} can be computed to be~\cite{Razamat:2013jxa},
\be
&&{\cal D}_{a,m}(\q,\t)\,\cdot\, f(a,m)=\\
&&\frac{1-\q^{m-\frac12}\t^{}a^{-2}}{1-\q^{m}a^{-2}}
\frac{1-\q^{m+\frac12}\t^{-1}a^{2}}{1-\q^{m}a^{2}}\;f(\q^{\frac12}\,a,m)+
\frac{1-\q^{m+\frac12}\t^{-1}a^{-2}}{1-\q^{m}a^{-2}}
\frac{1-\q^{m-\frac12}\t^{}a^{2}}{1-\q^{m}a^{2}}\;f(\q^{-\frac12}\,a,m)+\nonumber\\
&&\qquad\qquad\left(\frac{\t}{\q^{\frac12}}\right)
\frac{1-\q^{m+\frac12}\t^{-1}a^{-2}}{1-\q^{m}a^{-2}}
\frac{1-\q^{m+\frac12}\t^{-1}a^{2}}{1-\q^{m}a^{2}}\;f(a,m+1)+\nonumber\\
&&\qquad\qquad\qquad\left(\frac{\q^{\frac12}}{\t}\right)\,
\frac{1-\q^{m-\frac12}\t^{}a^{-2}}{1-\q^{m}a^{-2}}
\frac{1-\q^{m-\frac12}\t^{}a^{2}}{1-\q^{m}a^{2}}\;f(a,m-1)\,.\nonumber
\ee This operator acts by shifting the fugacity and the magnetic flux of a flavor symmetry.

In fact, in the $3d$ limit, this difference operator  can be factorized into two commuting operators:
\be
&&{\cal D}_{a,m}(\q,\t)\,\cdot\, f(a,m)={\cal O}_{(0,1)}{\cal O}_{(1,0)} \cdot f(z,\bar z)\,,
\ee
where
\be
\label{s2s1difdef}
{\cal O}_{(0,1)} f(a,m) = \t^{\frac{1}{2}} \q^{-\frac{1}{4}} \bigg(\frac{1 - \t^{-1} \q^{m+\frac{1}{2}} a^2}{1-\q^{m} a^2} f(\q^{\frac{1}{4}} a,m+\frac{1}{2}) + \frac{1 - \t^{-1} \q^{\frac{1}{2}-m} a^{-2}}{1-\q^{-m} a^{-2}}  f(\q^{-\frac{1}{4}} a,m-
\frac{1}{2})\bigg)\,, \nonumber \\
{\cal O}_{(1,0)} f(a,m) = \t^{-\frac{1}{2}} \q^{\frac{1}{4}} \bigg(\frac{1 - \t \q^{m-\frac{1}{2}} a^{-2}}{1-\q^m a^{-2}} f(\q^{\frac{1}{4}} a,m-\frac{1}{2}) + \frac{1 - \t \q^{-\frac{1}{2}-m} a^{2}}{1-\q^{-m} a^{2}}  f(\q^{-\frac{1}{4}} a,m+
\frac{1}{2})\bigg)\,. \nonumber \\ \nonumber 
\nonumber\\
\ee
Note that acting with only one of the two difference operators is physically ill defined
if the flavor group is $SU(2)$:
it involves shifting $m$ by half an integer
which is only allowed if the group is $SO(3)$.  

\

\subsection*{Holomorphic Blocks}

Let us now see how the $\S^3_b$ and $\S^2 \times \S^1$ difference operators, (\ref{s3difdef}) and (\ref{s2s1difdef}), both
descend from a single difference operator
acting on the  holomorphic blocks. Let us consider the $\S^3_b$ case first.  Then, with one caveat to be discussed below, we can rewrite (\ref{s3difdef}) as (in the notations of (\ref{s3blockdef}))
\begin{equation}
\label{diffo}
\mathcal{O}_{(0,1)} = \t^{\frac{1}{2}} \q^{-\frac{1}{4}} \bigg(\frac{1 - \t^{-1} \q^{\frac{1}{2}} \a^2}{1-\a^2} {p_{\a}}^{\frac{1}{2}}+ \frac{1 - \t^{-1} \q^{\frac{1}{2}} \a^{-2}}{1-\a^{-2}} {p_{\a}}^{-\frac{1}{2}} \bigg)
\end{equation}
Here $p_{\a}$ the operator which shifts ${\a} \rightarrow  \q {\a}$ while fixing $\tilde{\a}$, or equivalently, shifts $m \rightarrow m+ib$.  Since this expression also only depends on the untilded variables of (\ref{s3blockdef}),  we see that $\mathcal{O}_{(0,1)}$ acting on a factorized expression for $Z_{\S^3_b}$ as in (\ref{blockfact}) only acts on the left blocks, $B_\alpha(\z_a;\q)$, without modifying the right blocks $B_\alpha(\tilde{{\z}}_a;\tilde{\q})$.  One can similarly check that the operator $\mathcal{O}_{(1,0)}$ acts only on the right blocks.

The caveat mentioned above is that $p_{\a}^{\frac{1}{2}}$, which appears above, shifts $m \rightarrow m + \frac{i b}{2}$, and this also acts on the tilded variable $\tilde{\a}=e^{2 \pi b^{-1} m}$ by taking $\tilde{\a} \rightarrow -\tilde{\a}$.  However, we have seen in (\ref{bwt}) that we can choose blocks which are even under $\a \rightarrow -\a$, and then the difference operator indeed acts only on one set of blocks.

The basic difference operator $\mathcal{O}_{(0,1)}$ acting on the $\S^2 \times \S^1$ partition function,  (\ref{s2s1difdef}), can also be rewritten in terms of the variables (\ref{s2s1blockdef}).  For example, shifting $(a,m) \rightarrow (\q^{\frac{1}{4}} a,  m+\frac{1}{2})$ corresponds to taking ${\a} \rightarrow \q^{\frac{1}{2}}{\a}$ while not modifying $\bar{{\a}}$, {\it i.e.}, it is the operation $p_{\a}$.  One recovers exactly the same expression (\ref{diffo}).  In particular, it also only acts on the left blocks in the decomposition (\ref{blockfact}).  Thus we see that the difference operator acts naturally at the level of holomorphic blocks, and the same operator acting on the blocks gives rise, after fusion, to the $\S^3_b$ and $\S^2 \times \S^1$ difference operators we have constructed above.

\

\subsection*{$T[SU(2)]$ as an eigenfunction}

As we mentioned in the beginning of the section one can expect that the partition functions of
the $T[SU(2)]$ theory are eigenfunctions of the operators we reviewed above. This was verified 
for the $\S^3_b$ partition function
in \cite{Bullimore:2014nla}.  In the previous section we saw that these operators act at the level of the holomorphic blocks, so it is natural to ask how they act on the blocks of the $T[SU(2)]$ theory.  In fact, we claim that these blocks are eigenfunctions of the operators, with an eigenvalue which is independent of the block index $\alpha$.  In particular, this implies both the $\S^3_b$ and $\S^2 \times \S^1$ partition functions of this theory are eigenfunctions of both difference operators.

The argument will be an adaptation of the one appearing in \cite{Bullimore:2014nla} for the case of $\S^3_b$.  First it will be convenient to consider the theory we obtain before gauging the $U(1)_g$ gauge symmetry, which is just a theory of free hypers in the bifundamental representation of $U(1)_g \times SU(2)_V$.  This theory has a single block, which from (\ref{hyperblocks}) can be written as: 
\be B_{bif}({\t},{\z},{\a};\q) =  \frac{(\q^{\frac{3}{4}} \t^{-\frac{1}{2}} \z^{-1} \a^{\pm 1};\q)}{( \q^{\frac{1}{4}} \t^{\frac{1}{2}} \z^{-1} \a^{\pm 1};\q)}\,.\ee
Now let us consider the action of $\mathcal{O}_{(0,1)}$ on this expression.  First note that, for a single hypermultiplet, one has:
\be p_\z B_{hyp}({\t},{\z};\q) =  \frac{1-\q^{-\frac{1}{4}} \t^{-\frac{1}{2}} \z^{-1}}{1-\q^{-\frac{3}{4}} \t^{\frac{1}{2}} \z^{-1}}B_{hyp}({\t},{\z};\q)\,.
\ee
Thus, acting on the bifundamental hyper, one finds:
\be
&&{p_{\z}}^{\frac{1}{2}} {p_{\a}}^{\pm \frac{1}{2}} B_{fun}({\t},{\z},{\a};\q) = S_{\pm} B_{bif}({\t},{\z},{\a};\q)
\,,\\
&& p_{\z} B_{bif}({\t},{\z},{\a};\q) = S_+ S_- B_{fun}({\t},{\z},{\a};\q)\,,\nonumber
\ee
where (here it is convenient to define $\tilde{\x}=\q^{\frac{1}{2}} \t^{-1}$, analogous to the Coulomb limit variable $\tilde{x}$ in the index):
\be S_\pm = \frac{1-\q^{-\frac{1}{2}} \tilde{\x}^{\frac{1}{2}}  \z^{-1} \a^{\mp 1} }{1-\q^{-\frac{1}{2}} \tilde{\x}^{-\frac{1}{2}} \z^{-1} \a^{\mp 1}}\,.\ee
Meanwhile, from (\ref{diffo}), the difference operator is given by:
\be \mathcal{O}_{(0,1)} = T_+ {p_{\a}}^{\frac{1}{2}} +  T_- {p_{\a}}^{-\frac{1}{2}}\,,\ee
where
\be 
T_\pm = \tilde{\x}^{-\frac{1}{2}} \frac{1 - \tilde{\x} \a^{\pm 2}}{1-\a^{\pm 2}}\,.
\ee
Thus we can write
\be
\mathcal{O}_{(0,1)}B({\t},{\z},{\a};\q)  = {p_{\z}}^{-\frac{1}{2}} ( T_+ S_+ + T_- S_-) B({\t},{\z},{\a};\q)\,.
\ee
One checks that
\be T_+ S_+ + T_- S_- = \tilde{\x}^{-\frac{1}{2}} S_+ S_- +  \tilde{\x}^{\frac{1}{2}} \,,
\ee
so that 
\be
\mathcal{O}_{(0,1)}B({\t},{\z},{\a};\q)  =  {p_{\z}}^{-\frac{1}{2}}  ( \tilde{\x}^{-\frac{1}{2}}S_+ S_- + \tilde{\x}^{\frac{1}{2}}) B({\t},{\z},{\a};\q)  = ( \tilde{\x}^{-\frac{1}{2}} {p_{\z}}^{\frac{1}{2}} + \tilde{\x}^{\frac{1}{2}} {p_\z}^{-\frac{1}{2}} )B({\t},{\z},{\a};\q) \,. \nonumber \\
\ee
Thus the following relation holds in the algebra of line operators on the ungauged theory:
\be
\label{algrel} \mathcal{O}_{(0,1)} =\tilde{\x}^{-\frac{1}{2}} {p_{\z}}^{\frac{1}{2}} +\tilde{\x}^{\frac{1}{2}}  {p_\z}^{-\frac{1}{2}} \ee
Next we will go to the theory where the $U(1)$ symmetry corresponding to ${\z}$ is gauged, {\it i.e.}, the $T[SU(2)]$ theory.  Here one expects to find that the decomposition is no longer into a single left- and right- block, but rather into a sum of blocks, $B_\alpha$, as in (\ref{blockfact}).  As argued in \cite{Beem:2012mb}, this gauging is accomplished at the level of the algebra of line operators by introducing operators $\b,p_\b$ for the new $U(1)_J$ symmetry, and making the following replacement in (\ref{algrel}):\footnote{Here there are extra factors of $2$ in these assignments because we have normalized the fugacity $\b$ to couple to the $U(1)_J$ symmetry with charge $2$.}
\be {\z} \rightarrow p_{\b'}^{\frac{1}{2}}, \;\;\; p_{\z} \rightarrow {\b'}^{-2}\,,\ee
so that:
\be\mathcal{O}_{(0,1)}B_\alpha^{T[SU(2)]} ({\t},{\a},{\b};\q)  = ( \tilde{\x}^{-\frac{1}{2}} {\b}' + \tilde{\x}^{\frac{1}{2}}{{\b}'}^{-1} )B_\alpha^{T[SU(2)]}({\t},{\a},{\b};\q) \,.\ee
Finally, we recall that the partition function of the hypermultiplet built from the blocks contained a background FI term, so that $\b'$, which corresponds to the bare FI term, must be shifted to obtain the full FI term, which corresponds to the variable $\b$ above.  One can check that this shift is by precisely $ \tilde{\x}^{\frac{1}{2}}$, and cancels the factors on the terms above.  Thus we arrive at the final result:
\be\mathcal{O}_{(0,1)}B_\alpha^{T[SU(2)]} ({\t},{\a},{\b};\q)  = ({\b} +{{\b}}^{-1} )B_\alpha^{T[SU(2)]}({\t},{\a},{\b};\q) \,,\ee
which is precisely the contribution of a Wilson loop in the fundamental representation of the $SU(2)$ flavor symmetry corresponding to ${\b}$.

Since we have an explicit expression for the blocks of $T[SU(2)]$, given in (\ref{bwt}), by expanding this series to high order we can explicitly verify that this is indeed an eigenfunction of the difference operator (\ref{diffo}).  Here it is important that we use the blocks which are symmetric under $\a \rightarrow -\a$ so that the difference operator acts only on one set of blocks.  

When we think of $T[SU(2)]$ as an $S$-duality wall for $\mathcal{N}=4$ SYM, the relation \eqref{algrel} can be interpreted as the equivalence of the basic 't Hooft loop in one $su(2)$ factor with the basic Wilson loop of the other \cite{Gang:2012ff,Bullimore:2014nla}. The choice of blocks with the $\a \rightarrow -\a$ symmetry has a natural interpretation in this context - it amounts to choosing the Higgs flavor symmetry of $T[SU(2)]$ to be $SO(3)$ rather than $SU(2)$, so that it becomes well-defined to act with an 't Hooft loop operator.  Then the Coulomb flavor symmetry, parameterized by $\b$, is $SU(2)$, and one cannot consistently act with such an operator.  These roles are exchanged under mirror symmetry.

\

When one thinks of the partition functions, $\psi_{\q,\t}(a,m|b,n)$ and $\phi_\gamma(a|b)$,
as eigenfunctions of the difference operators one set of variables, say $(a,m)$ for the index
and $a$ for the $\S^3$ partition function, label the spectrum. One can think then of this set of variables
as the ``momenta'' and the other as the ``position'' of the particles on the circle with the Hamiltonians
being the difference operators. Then mirror symmetry is the duality exchanging momenta and positions
in this language. In fact this kind of a duality of integrable model and its relations to different
gauge theories has been discussed already a while ago, see {\it e.g.}~\cite{Fock:1999ae,Gorsky:2000px}.

\

We should stress that although the $3d$ difference operators one obtains are Macdonald 
operators, the eigenfunctions relevant for the $3d$ partition functions are not Macdonald
polynomials. This is because the measure under which we expect the eigenfunctions to be
orthogonal, the vector multiplet measure, is not the Macdonald measure. The orthogonality property 
for the $\S^3_b$ partition functions in the Schur limit appears for
 example in~\eqref{ortho}. 
 The difference operators 
in $4d$ are of  elliptic type (elliptic in ``positions'' and trigonometric in ``momenta'') and their 
spectrum is hard to obtain in closed form (see {\it e.g.}~\cite{Razamat:2013qfa}). However, in $3d$
since we have a physical meaning of the eigenfunctions as partition functions of theories with known
Lagrangian description the computation of the eigenfunctions is straightforward.

\

\subsection{Poles}

The partition functions of $T[SU(2)]$ has yet another interesting property which it inherits from $4d$. 
The difference operators above are derived by computing certain residues of the index
as reviewed in section~\ref{sec4d}. 
Correspondingly the residues of the partition functions of $T[SU(2)]$ give the eigenvalues of the
difference operators, see~\eqref{residuesQ}.
Let us compute several residues of the $\S^2\times\S^1$ partition function of $T[SU(2)]$.

\

\noindent {\bf{ 
Residue at $(a=\t^{\frac12}\,\q^{\frac14},\;m=0)$}}

\

This is the basic residue: the pole is obtained from the $\widetilde m=0$ sector when 
two poles pinch the integration contour and we readily get,
\be
{\cal I}_V\;Res_{a\to\t^{\frac12}\,\q^{\frac14}}\;\psi_{\q,\t^{}}(b,n|a,0) = 1\,.
\ee  Here ${\cal I}_V$ is the index of the vector multiplet.

\

\noindent {\bf{ 
Residue at $(a=\t^{\frac12}\,\q^{\frac34},\;m=0)$}}

\

This is the first non-trivial residue when the flavor group is taken to be $SU(2)$: the pole is obtained from the $\widetilde m=0,\pm1$ sector when 
two poles pinch the integration contour (for $\widetilde m=0$ the poles pinch at $z=\q^{\pm\frac12}$ 
and for $\widetilde m=\pm1$ they pinch at $z=1$) and we readily get,
\be
{\cal I}_V\;\frac{\t}{\q^{\frac12}}\left(\frac{1-\q}{1-\t\q^{\frac12}}\right)^2\;
Res_{a\to\t^{\frac12}\,\q^{\frac34}}\;\psi_{\q,\t^{}}(b,n|a,0) = (b \q^{\frac{n}{2}}+b^{-1} \q^{-\frac{n}{2}})(b \q^{-\frac{n}{2}}+b^{-1} \q^{\frac{n}{2}})\,.
\ee Note that this corresponds to the action of the difference operator introducing a pair of line
defects and the residue is given just by a product of partition functions of two Wilson lines.

\

\noindent {\bf{ 
Residue at $(a=\t^{\frac12}\,\q^{\frac12},\;m=\pm\frac12)$}}

\

This is the first non-trivial residue when the flavor group is taken to be $SO(3)$: the pole is obtained from the $\widetilde m=\pm\frac12$ sector when 
two poles pinch the integration contour (at $z=q^{\pm\frac14}$) and we readily get,
\be
{\cal I}_V\;\frac{\t^{\frac12}}{\q^{\frac12}}\left(\frac{1-\q}{1-\t\q^{\frac12}}\right)\;
Res_{a\to\t^{\frac12}\,\q^{\frac34}}\;\psi_{\q,\t^{}}(b,n|a,\pm\frac12) = 
b \q^{\pm\frac{n}{2}}+b^{-1} \q^{\mp\frac{n}{2}}\,.
\ee Here the residues are just single Wilson lines. 

\

\noindent In general the residues of indices are expected to describe indices of IR fixed points
reached by turning on vacuum expectation values for certain operators~\cite{Gaiotto:2012xa}.
 The residues obtained above thus correspond to empty (free)
theories with line operators for non-dynamical gauge fields.

\

\section{$\dots$ and back again: $4d$ shards of $3d$ mirrors}

In this final section we will  make some speculative remarks about what  the structure one finds for the $3d$ reductions
of theories of class ${\cal S}$ implies about the $4d$ theories. Let us start from the trivial case
of a partition function capturing physics which is invariant under the dimensional reduction.

\subsection*{Hall-Littlewood/Higgs limit}

An example of such a partition function is the  Hall-Littlewood (HL) index in $4d$~\cite{Gadde:2011uv}.
For theories of class ${\cal S}$ corresponding to genus zero Riemann surfaces  this index is equivalent to the Hilbert series of the Higgs branch.  Note from (\ref{lensredef}) that, in the Hall-Littlewood limit, $p=q=0$, the lens index is independent of $r$, provided the dimensionally reduced $3d$ theory is ``good/ugly''.\footnote{This condition is required so that the zero point energy contains only positive powers of $p$ and $q$, and so is well-defined in this limit.}  In the three dimensional limit this index reduces to the Higgs index defined in section $3$, and so for such theories the Hall-Littlewood index of the four dimensional parent theory matches with the Higgs index of its three dimensional reduction.  The latter in turn is equal to the Coulomb index of the mirror dual. Let us discuss
the HL index of the $T_2$ theory in the mirror, star-shaped, frame.  This index is 
given in~\eqref{TSU2p} and the indices of the legs are evaluated in~\eqref{legHL}. Putting
these ingredients together we obtain in the HL limit that,

\be
{\cal I}^H_{T_2}(\{[b_i,0]\}_{i=1}^3)_{free}&=&
\sum_{\widetilde m\in  \Z/2} \,
\frac{x^{-2|\widetilde m|}}
{1-x}\, \oint\frac{dz}{4\pi i z}
\left(\frac{1-z^{\pm2}}
{1-x z^{\pm2}}\right)^{\delta_{\widetilde m,0}}\prod_{\ell=1}^3
\psi_{x}(b_\ell,0|z,\widetilde m)\,\nonumber\\
&=&
\sum_{\widetilde m\in  {\mathbb N}/2} \,
\frac{\left(x \right)^{|\widetilde m|}(1+x)^{-\delta_{\tilde m,0}}}{1-x}
\prod_{i=1}^3
\frac{1}{1-x\,b_i^{\pm2}}\; \chi^{(HL)}_{|2\widetilde m|}(b_i;x)\,.
 \ee
This is precisely the HL index of the $T_2$ theories written in the form of~\eqref{maineq}.
Thus, for the HL index the eigenfunctions appearing in $4d$ have a concrete $3d$ 
physical meaning: they are the Coulomb indices of the $T[SU(2)]$ theories.  A similar observation was made in \cite{Cremonesi:2014vla} from the point of view of the Coulomb branch Hilbert series, which we have claimed is equivalent to the Coulomb limit.
The fugacity of the $SU(2)$ flavor symmetry acting on the Coulomb branch in $3d$ is the flavor 
fugacity in $4d$, and the label of the eigenfunction is the background monopole charge for 
the symmetry acting on the Higgs branch. This observation has a straightforward generalization
to higher rank cases: HL eigenfunctions are the Coulomb indices of $T[SU(N)]$ theories.
The discrete labels of the eigenfunctions
are the GNO charges of the monopole background for the Coulomb branch flavor symmetry.
We can write the Coulomb index of the $T[SU(N)]$ theory as
\be\label{HLAn}
&&\psi_{\lambda_1,\lambda_2,\dots}({\bf z}^{(N)};0,0,t)=
\prod_{i=1}^{N-1}\left[\frac{(1-t)^{i}}{i!}
\oint \prod_{j=1}^i\frac{\left(z_j^{(i)}\right)^{n_i}dz_j^{(i)}}{2\pi i z_j^{(i)}}
\prod_{j_1\neq j_2}^i(1-tz^{(i)}_{j_1}/z^{(i)}_{j_2})(1-z^{(i)}_{j_1}/z^{(i)}_{j_2})
\right]\times\nonumber\\
&&\qquad\qquad\prod_{i=1}^{N-1}\prod_{j_1=1}^i
\prod_{j_2=1}^{i+1}{\cal I}_H((z_{j_1}^{(i)})^{-1}z_{j_2}^{(i+1)};0,0,t)\,.
\ee Here we have
\be
\lambda_i=\sum_{j=1}^{N-i} n_{N-j}\,.
\ee The functions ${\cal I}_H(z;0,0,t)$
is the HL index of a $4d$ hypermultiplet,
\be
{\cal I}_{hyp}^{4d}(z;p,q,t)=\prod_{i,j=0}^\infty\prod_{s=\pm1}\frac{1-\frac{p\,q}{t^{\frac12}} p^i q^j z^{s}}
{1-t^{\frac{1}{2}} p^iq^j z^{s}}\equiv
\Gamma(t^{\frac12} z^{-1}\, ;p,q)
\Gamma(t^{\frac12} z\, ;p,q)\,.
\ee

\

We can now construct the HL index of any theory of class ${\cal S}$ from the star-shaped quivers.
To build an index of a general linear quiver we can glue together the star shaped 
mirrors of the free bi-fundamental hypermultiplets by gauging $SU(2)$ global
symmetries acting on the Coulomb branch. One does so by using the usual vector multiplet.
Increasing the genus can be done in two ways. First we can gauge a diagonal combination
of two $SU(2)$s acting on Coulomb branches of two different legs.
  This procedure is manifestly 
equivalent to the $4d$ procedure and the result is

\be\label{gluegenus}
{\cal I}^H(\{[b_i,0]\}_{i=1}^s)^A_{g,s}&=&
\sum_{\widetilde m\in  {\mathbb N}/2} \,
\frac{x^{|\widetilde m|(2g-2+s)}(1+x)^{(g-1)\delta_{\tilde m,0}}}{(1-x)^{1-g}}
\prod_{i=1}^s
\frac{1}{1-x\,b_i^{\pm2}}\; \chi^{(HL)}_{|2\widetilde m|}(b_i;x)\,.\nonumber\\
 \ee
A second procedure one can try is as follows. Since the mirror
of a genus $g$ theory of class ${\cal S}$ is the same as the one for genus zero with an addition
of $g$ adjoint (under the $SO(3)$ of the central node) hypermultiplets~\cite{Benini:2010uu},
 we can just add the contribution
of these fields to the index computation. The adjoint hypermultiplets contribute a factor 
of $\left(\t\q^{\frac12}\right)^{2|\widetilde m|\,g}$. The resulting index is 

\be{\cal I}^H(\{[b_i,0]\}_{i=1}^s)^B_{g,s}&=&
\sum_{\widetilde m\in  {\mathbb N}/2} \,
\frac{x^{|\widetilde m|(2g-2+s)}(1+x)^{-\delta_{\tilde m,0}}}{1-x}
\prod_{i=1}^s
\frac{1}{1-x\,b_i^{\pm2}}\; \chi^{(HL)}_{|2\widetilde m|}(b_i;x)\,,\nonumber\\
 \ee and is manifestly different from~\eqref{gluegenus} for $g\neq 0$.
There are two reasons why the two procedures disagree: one is  that
 the Hall-Littlewood/Higgs index of a higher
genus quiver is not the same as the Hilbert series of the Higgs branch; and
the second one is that the higher genus quivers are {\it ``bad''} theories in $3d$.
This implies that for higher genus theories the first procedure does not make physical sense
in $3d$: the index of a bad theory is divergent\footnote{
A dimensional reduction of $4d$ theories with perfectly finite partition functions
might produce $3d$ theories with divergent partition functions. For example
this might happen if  the $3d$ monopole operators have zero charges, see {\it e.g.}~\cite{Aharony:2013kma}.} and by taking the Higgs limit
we obtained certain regularization which is not physically motivated. 
On the other hand, the mirror with extra adjoints is a good theory with finite index
and well defined Coulomb limit. To summarize: procedure A gives the $4d$ index but 
has no obvious meaning in $3d$, while procedure B gives the Coulomb index of the 
$3d$.  Procedure B seems however to be equivalent 
to the Hilbert series of the Higgs branch of the $4d$ ``parent'' theory as defined in~\cite{Hanany:2010qu}, see~\cite{Cremonesi:2014vla}.  The $3d$ 
Coulomb indices of the mirror duals of class $\frak s$ theories are equivalent to the
Hilbert series of the Higgs branches of the $4d$ theories of class ${\cal S}$. This 
statement is not in tension  with the fact that for higher genus theories the $4d$ 
HL index is not equivalent to the Hilbert series since the dimensional reduction of such 
models produces ``bad'' theories in $3d$. If a theory is ``good'' or ``ugly'' then the Higgs
index is equal to the Hilbert series of the Higgs branch and the Coulomb index is equal to the Hilbert
series of the Coulomb branch.

\

\subsection*{Going beyond Higgs branch}

If we are interested in quantities capturing properties of a theory beyond its Higgs branch
the relation between special structures found in $3d$ and the $4d$ origin becomes less clear.
However, we will discuss now an encouraging mathematical fact: the 
$4d$ eigenfunctions of the $\S^3\times \S^1$ partition function,
 at least in the Macdonald limit, can be written using integral expressions
with the same structure as the $3d$ partition functions of the $T[SU(N)]$ theories.

We discuss the Macdonald index in $4d$, $p=0$ and $r=1$.
The relevant eigenfunctions can be found to be closely related to Macdonald polynomials~\cite{Gadde:2011uv}. There is a very useful representation of Macdonald
polynomials using q-integrals~\cite{qinteg,macNest,qintegOK}.
In the $A_1$ case this takes the following form~\cite{qinteg} %\qinteg,

\be
P_n(b;q,t)=\frac{(b-b^{-1})(tb^{\pm2};q)}{(1-q)(b^{\pm1};q)}
\frac{(t;q)^2}{(t^2;q)(q;q)}\frac{(t^2;q)_n}{(q;q)_n}\;
\int_b^{b^{-1}}\frac{(qub^{\pm1};q)}{(tub^{\pm1};q)}\;u^n\;d_qu\,.
\ee The q-integrals are defined as

\be
&&\int_0^bf(u)d_qu = b(1-q)\sum_{k=0}^\infty f(bq^k)q^k\,,\\
&&\int_a^bf(u)d_qu = \int_0^bf(u)d_qu-\int_0^af(u)d_qu\,.\nonumber
\ee One can view the q-integral expression for the eigenfunction 
as a $4d$ quantity associated to $3d$ $T[SU(2)]$ theory: the $u^n$ term is  the contribution
of the FI term and the rest of the integral is a q-deformed contribution of the hypermultiplet.

We can actually write the q-integral above as a usual contour integral. 
We look for an expression of the form,

\be
&&\psi_n(a;0;q,t)=\frac{(t^2;q)}{(t;q)(t\,a^{\pm2};q)}\frac{P_n(a;q,t)}{P_n(t^{\frac12};q,t)}=
\\
&&\qquad\Phi(a;0,q,t)
(q;q)(t;q)\,\oint \frac{dz}{2\pi i z}\;z^{n}\;\Psi(z;0,q,t)
{\cal I}_{H,4d}(a,z;0,q,t)\,.\nonumber
\ee
Assuming $\Psi(z)$ does not have poles inside the unit circle, the sum of the poles of  this 
integral is the same as the sum in the definition of the q-integral for the Macdonald polynomial above given that,

\be
\Phi(z)=\Psi^{-1}(t^{1/2}z)\,,\qquad
\Psi(q\,z)=\frac{t}qz^{-2}\Psi(z)\,.
\ee This has the following solution
\be
&&\Psi(z)=\prod_{\ell=1}^\infty \left(1+q^{2\ell-1}\frac{z^2}{t}\right)\left(1+q^{2\ell-1}\frac{t}{z^2}\right)=
\theta(-\frac{q}{t}\,z^{2};q^2)\\
&&\qquad\qquad =\theta(i\,\sqrt{\frac{q}{t}}z;q)\;\theta(-i\,\sqrt{\frac{q}{t}}z;q)\,.\nonumber
\ee Note that in HL limit, $q=0$,  $\Psi(z)=1$ and we reproduce the result in the beginning of this section.

One can generalize the q-integral construction for $A_1$ to higher rank group 
and the result has a structure of q-deformation of the result for the HL case~\eqref{HLAn},
see~\cite{macNest,qintegOK}. 
It is thus interesting to understand whether there is any physical meaning of the q-integral
expression for the eigenfunctions  and/or of the contour integral one with $\Psi(z)$.

\

\section*{Acknowledgments}

We would like to thank C.~Beem, T.~Dimofte, A.~Hanany, and 
Y.~Tachikawa for useful discussions.  
 SSR gratefully acknowledges support from the Martin~A.~Chooljian and Helen Chooljian membership
 at the Institute for Advanced Study. The research of SSR was also partially supported by
NSF grant number PHY-0969448.  The research of BW was supported in part by DOE Grant DE-SC0009988. SSR and BW would like to thank KITP, Santa Barbara, and the Simons Center, Stony Brook, for hospitality and support during different stages of this work.

\bibliography{n43dpaper}
\bibliographystyle{JHEP}

\end{document}